\documentclass{aa} 
\usepackage{multirow}

\usepackage{graphicx}
 \usepackage[varg]{txfonts}
\usepackage[colorlinks=true,citecolor=blue,linkcolor=blue]{hyperref}
\usepackage[dvipsnames]{xcolor}
\usepackage[all]{hypcap}

\usepackage{lineno}
\linenumbers

\graphicspath{{./Figures/}}
\newcommand{\nustar}{\textsl{NuSTAR}\xspace}
\newcommand{\xmmnewton}{\textsl{XMM-Newton}\xspace}
\newcommand{\integral}{\textsl{INTEGRAL}\xspace}
\newcommand{\batse}{\textsl{BATSE}\xspace}

\DeclareUnicodeCharacter{2212}{-}
\DeclareUnicodeCharacter{2060}{\nolinebreak}

\begin{document}

\title{Pulse profile variations in the accreting X-ray pulsar Vela X-1}

\author{
V.~Madurga-Favieres\inst{\ref{affil:CSST},\ref{affil:GSFC},\ref{affil:CRESST}}\and
A.~Martin-Carrillo\inst{\ref{affil:UCD}}\and
C.~M.~Diez\inst{\ref{affil:ESAC}}\and
F.~Fürst\inst{\ref{affil:ESAC}}\and
S.~Mart\'inez-N\'u\~nez\inst{\ref{affil:IFCA}}\and
V.~Grinberg\inst{\ref{affil:ESTEC}}\and 
L.~Abalo\inst{\ref{affil:Leiden}}\and
I.~El~Mellah\inst{\ref{affil:USC},\ref{affil:USACH}}\and
P.~Tzanavaris\inst{\ref{affil:CSST},\ref{affil:GSFC},\ref{affil:CRESST},\ref{affil:APS}}\and
P.~Kretschmar\inst{\ref{affil:ESAC}}
}

\institute{%
% Vicente
Center for Space Science and Technology, University of Maryland, Baltimore County, 1000 Hilltop Circle, Baltimore, MD 21250, USA \label{affil:CSST} \email{vmadurga@umbc.edu ; vmadurgafavieres20@gmail.com} 
\and
X-ray Astrophysics Laboratory, ⁠NASA/Goddard Space Flight Center, Greenbelt, MD 20771, USA \label{affil:GSFC}
\and
Center for Research and Exploration in Space Science and Technology, NASA/Goddard Space Flight Center, Greenbelt, MD 20771,
USA \label{affil:CRESST}
\and %Antonio
School of Physics and Centre for Space Research, University College Dublin, Belfield, Dublin 4, Ireland \label{affil:UCD}
\and  %Peter
European Space Agency (ESA), European Space Astronomy Centre (ESAC), Camino Bajo del Castillo s/n, 28692 Villanueva de la Cañada, Madrid, Spain  \label{affil:ESAC}
\and %Silvia
Instituto de F\'isica de Cantabria (CSIC-Universidad de Cantabria), E-39005, Santander, Spain \label{affil:IFCA}
\and
European Space Agency (ESA), European Space Research and Technology Centre (ESTEC), Keplerlaan 1, 2201 AZ Noordwijk, The Netherlands \label{affil:ESTEC}
\and
Huygens-Kamerlingh Onnes Laboratory, Leiden University, Postbus 9504, 2300 RA Leiden, The Netherlands \label{affil:Leiden}
\and
Departamento de Física, Universidad de Santiago de Chile, Av. Victor Jara 3659, Santiago, Chile \label{affil:USC}
\and
Center for Interdisciplinary Research in Astrophysics and Space Exploration (CIRAS), USACH, Santiago, Chile \label{affil:USACH}
\and
American Physical Society, Hauppauge, New York, NY 11788, USA \label{affil:APS}
}

\date{Received: -- / accepted: --}

\abstract
  % context heading (optional)
    {\object{Vela X-1} is a well-studied accreting X-ray pulsar, with a distinctive pulse profile that has been found to be very similar in different observations spread out over decades. On the other hand, significant variations down to the timescale of individual pulses have been observed. The physical mechanisms leading to the energy-resolved pulse profile and its variations are not fully understood. 
   Long, uninterrupted observations of Vela X-1 with \xmmnewton in 2000, 2006 and 2019 at different orbital phases allow us to study variations of the pulse properties in the soft X-ray range on all timescales in detail.}
  % aims heading (mandatory)
   {We aim to characterize and quantify the variations of pulse profiles and individual pulse cycles on all timescales probed, and to identify possible factors driving the observed variations on these timescales.}
  % methods heading (mandatory)
   {We generated consistent energy-resolved pulse profiles for each observation, as well as profiles built from subsets of individual pulse cycles selected by time, flux, or similarity to the mean profiles. We identified five pulsed phases based on the profile morphology and hardness, and examined the relative contributions over time. To quantify short-timescale variability, we compared individual pulse cycles with synthetic light curves derived from scaled versions of the average profiles. }
  % results heading (mandatory)
   {The pulse profile of Vela X-1, when averaged over many pulse cycles, remains remarkably stable, as expected. The most prominent variations between epochs are attributable to changes in absorption. Residual systematic differences are primarily flux-dependent, with profiles showing less variability at higher flux levels. On shorter timescales, most individual pulse cycles resemble the average profile, even though significant, sporadic deviations are also present.}
   {}
   

   \maketitle

\section{Introduction}
\label{sec:introduction}

\begin{table*}
    \renewcommand{\arraystretch}{1.1}
    \caption{Overview of the three \xmmnewton observations of Vela X-1 with public data in the archive as of late 2024.}
    \centering
    \begin{tabular}{rrrrr@{--}lr}
    \hline\hline
    \multicolumn{1}{c}
    {Obs} & 
    \multicolumn{1}{c}{ObsID} &  
    \multicolumn{1}{c}{Dates} &
    \multicolumn{1}{c}{Duration (s)} &
    \multicolumn{2}{c}{MJDs} &
    \multicolumn{1}{c}{Orbital phase} 
    \\
    \hline
      1 
      & 
      0111030101 
      & 
      2000 Nov 02-03 
      & 
      59114 
      &
      51850.580 
      & 
      51851.264
      & 
      0.654--0.730  \\  
      2 
      & 
      0406430201 
      & 
      2006 May 25-26 
      & 
      124310
      &
      53880.438 
      & 
      53881.877
      & 
      0.090--0.251 \\
      3 
      & 
      0841890201 
      & 
      2019 May 03-05 
      & 
      114900
      &
      58606.885 
      & 
      58608.215
      & 
      0.339--0.488
      \\
    \hline
    \end{tabular}
    \tablefoot{The orbital phases were obtained with the ephemeris in Table 4 from \citet{Kreykenbohm+2008}, using $P_{\textrm{orb}} = 8.964357 \pm 0.000029$~d and $T_{\textrm{ecl}}$ as time of phase zero.}
    \label{tab:observations table}
\end{table*}

\object{Vela X-1} is one of the best-studied high-mass X-ray binary pulsars, since it was detected early \citep{Chodil+1967}, shows persistent X-rays with clear pulsations \citep{McClintock+1976}, and has a rich phenomenology in a wide range of wavelengths. 
The following parameters are taken from the recent review by \citet{Kretschmar+2021}. 
The system lies at a distance of $\sim 2$~kpc with an intrinsic luminosity of $\sim 4\times10^{36} \mathrm{erg} \mathrm{s}^{-1}$. 
It is formed by the B0.5 Ia supergiant HD77581 ($M_{\star} \approx 20 - 30 M_{\odot}$, $R_{\star} \approx 30R_{\odot}$) and a neutron star ($M_{NS} \approx 1.7 - 2.1M_{\odot}$, $R_{NS} \approx 11 - 12$~km). 
The neutron star orbits the supergiant in an almost circular orbit ($e \approx 0.0898$) with a period of $\sim 8.964$~days and a mean distance of $\sim 1.7 R_{\star}$. X-ray eclipses of $\sim$20\% of the orbit indicate that the orbital plane is observed near edge-on.

Being deeply embedded in the acceleration zone of the supergiant's strong stellar wind \citep{Martinez+2014, Kretschmar+2021}, the X-ray emission from the neutron star is significantly affected by photoelectric absorption at all orbital phases. In general, the observed absorption clearly decreases at eclipse egress and then rises again to higher values around orbital phase 0.5 (where mid-eclipse is phase zero), but with strong variations around this mean pattern \citep{Sato:1986PASJ, Doroshenko:2013, Kretschmar+2021, Abalo+2024}.

The X-ray spectrum of Vela X-1 is similar to that of other accreting X-ray pulsars and has been mostly described by a powerlaw modified by variable strong photoelectric absorption at lower energies and an exponential cutoff beyond $\sim 20$\,keV, further modified by two cyclotron lines at  $\sim 25$\,keV and $\sim 54$\,keV \citep{Furst+2014}. In addition, a strong iron fluorescence emission line is present.

\citet{Fuerst:2010} carried out a study of hard X-ray data from \textsl{INTEGRAL}. They found a log-normal distribution of observed fluxes, with a median absolute luminosity $\langle L_\mathrm{X}\rangle = 5.1 \times 10^{36}$\,erg\,s$^{-1}$ and multiplicative standard deviation $\sigma\approx 2$.

Vela X-1 pulsates with a period of $\sim 283$~s. The period varies on all timescales, from days to years, described by a random walk in pulse frequency \citep{Boynton+1984, Deeter+1989}. Within the history of pulse period determinations, the range of variability is about $\pm 0.5$~s \citep[][Fig.~8]{Kretschmar+2021}.

The broadband pulse profile shows two broad pulses (see, e.g., Fig.~1 in \citet{Staubert+1980} and Sect.~\ref{sec:energy-resolved pulse profiles and hardness ratios}), which have been related to the contributions from emission regions near the two magnetic poles \citep{McClintock+1976, Kraus+1986}, similar to many other accreting pulsars.
The overall pulse profile shape has been found to be very stable over decades -- compare, e.g., the profiles in \citet{McClintock+1976, Staubert+1980, Raubenheimer+1990, Barbera+2003}; or \citet{Alonso+2022}. The pulse profile is strongly energy dependent. Below 6\,keV the shape is complex, with up to five visible peaks. At higher energies ($\geq 10$\,keV) it evolves into a double-peaked profile. The relative strength of these two peaks also changes with energy \citep{Raubenheimer+1990, Barbera+2003}. 

Variations of the X-ray emission in accreting X-ray pulsars from one pulse cycle to the next are commonly observed but relatively rarely studied more systematically. For \object{Vela X-1}, \citet{Staubert+1980} compared individual pulse cycles and the mean pulse profile during a hard X-ray balloon observation, finding significant pulse-to-pulse variations, but without distinguishing between overall flux changes and shape variability. Using Suzaku/PIN data, \citet{Kretschmar+2014} applied a rolling scale factor to the mean profile before the comparison to allow for flux changes. They found deviations between the scaled profile and the actual light curve on the order of a few 10\%, but also regular structures in the residuals, and did not study changes in the shape of single pulse cycles as function of flux. Variations at short timescales have also been reported for other sources, e.g., \object{1A 0535+262} \citep{Frontera+1985, Klochkov+2011}, \object{Cen~X-3} \citep{Mueller+2011}, \object{GX~301$-$2} \citep{Furst+2011}, or \object{4U 0114+65} \citep{Sanjurjo-Ferrin:2025}.

Over the years, there have been multiple attempts to model the pulsed emission from Vela~X-1, like for other accreting X-ray pulsars \citep[e.g.,][]{Sturner+Dermer:1994, Bulik+1995, Leahy+Li:1995, Laycock:2025}. But these different studies have used different assumptions and approximations and have arrived at very different solutions for the emission geometry. We, therefore, still lack a reliable base to explain the pulse profiles and their variations based on an emission model. Thus, we focus on an empirical study, which can be later used, for example, as benchmark to test emission models against.

In the following, we present a detailed analysis of pulse profile and pulse-to-pulse variations of Vela X-1 in the 1--10\,keV range, based on three long observations of the source with \xmmnewton -- see Sect.~\ref{sec:observations and data reduction} below. Section \ref{sec:light curves and pulse profiles per observation} presents the energy-resolved light curves and pulse profiles obtained from these observations. In Sect.~\ref{sec:pulse profile variations with time and flux}, we discuss the time- and flux-dependence of the pulse profiles. We analyze variations of individual pulse cycles in  Sect.~\ref{sec:variations of individual pulse cycles}. The overall results are discussed in Sect.~\ref{sec:discussion}, and a summary and future outlook are given in Sect.~\ref{sec:summary and outlook}. 

\section{Observations and data reduction}
\label{sec:observations and data reduction}

\subsection{Observations}
\label{sec:observations}

\begin{figure}
    \centering
    \includegraphics[width=0.5\textwidth]{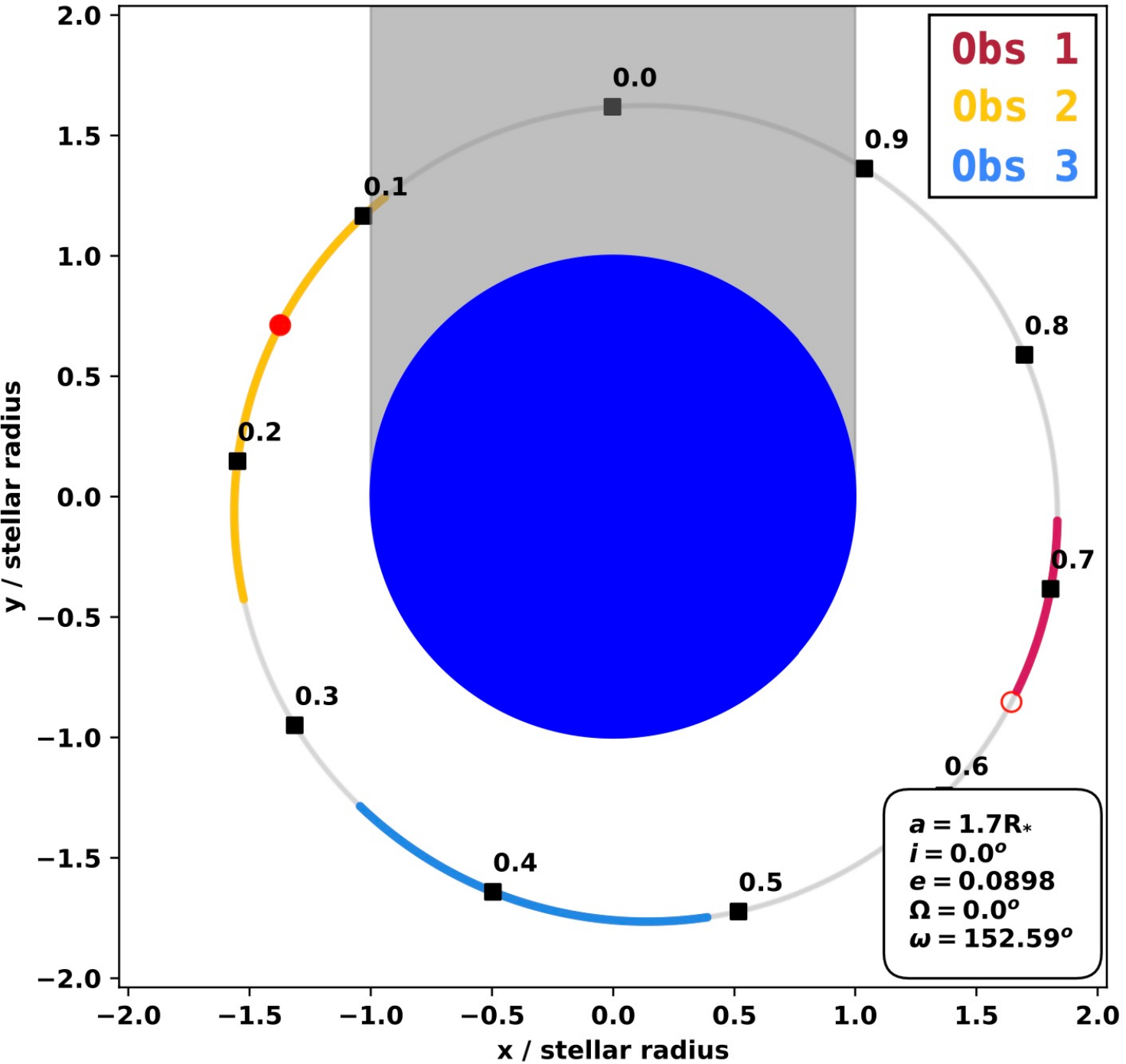}
    \caption{Top view of the Vela X-1 system during the three \xmmnewton observations analyzed in this article. The red circles represent the periastron (filled) and the apastron (empty). The observer is located facing the system at $x = 0$ and at minus infinity along the y-axis. The gray zone represents the eclipse range.} 
    \label{fig:orbit and observations}
\end{figure}

In this study, we analyzed the only three \xmmnewton observations of \object{Vela X-1} with public data available in the archive as of late 2024, referred to hereafter as Obs~1, Obs~2 and Obs~3, respectively. In Table~\ref{tab:observations table} and Fig.~\ref{fig:orbit and observations}, we provide an overview of these observations and the orbital ranges they covered.

For Obs~1, at a later phase of the orbit, no detailed study has been published so far. Obs~2, studying the egress from eclipse, has been studied by \citet{Martinez+2014} and \citet{Lomaeva+2020}. Obs~3 successfully aimed at observing the accretion wake moving into the line of sight \citep{Diez+2023}.

Slightly different combinations of the X-ray instruments were used in each observation. In this publication we focus exclusively on data from the European Photon Imaging Camera pn-CCDs \citep[EPIC-pn;][]{Strueder+2001} in timing mode, due to its high temporal resolution and ability to mostly avoid saturation and mitigate pile-up for this bright X-ray source. 

\subsection{Data reduction}
\label{sec:data reduction}

To generate the event lists, we used the Science Analysis System (SAS) software v21.0 with the Current Calibration Files as of March 2024, starting from the Observation Data Files level, running \texttt{epproc} with default calibration. Although \citet{Diez+2023} proposed a method in their Appendix to correct for an offset of instrumental and physical spectral lines toward high energies observed for Vela X-1 with EPIC-pn in timing mode, we extracted the events with the default Rate-Dependent PHA calibration, as this work focuses on timing analysis. We checked for flaring particle background in all observations, but no filtering was needed. In addition, we did not extract background for any of the observations, as the source was so bright that it illuminated the entire CCD plane. The event times were barycentered and deleted if found on bad pixels. We removed the outermost part of the point spread function wings to reduce the influence of background and possible dust scattering effects, leading us to extract events in the source region \texttt{RAWX} in [34:40] for Obs~1, \texttt{RAWX} in [31:43] for Obs~2, and \texttt{RAWX} in [32:44] for Obs~3. 

The arrival times were corrected for the binary orbit. We specifically used the \texttt{BinaryCor} function from the Interactive Spectral Interpretation System (ISIS) v1.6.2-51 \citep{Houck+2000} scripts (\texttt{ISISscripts}), provided by ECAP/Remeis Observatory and MIT\footnote{\url{http://www.sternwarte.uni-erlangen.de/isis/}}, and the ephemeris from \citet{Kreykenbohm+2008}.

All subsequent analysis was restricted to the 1--10\,keV range, due to very low count rates below 1\,keV and the relatively limited effective area of the EPIC-pn above 10\,keV (see Figs.~11 and~12 in Sect. 3.2.2.1 of the \xmmnewton Users Handbook\footnote{\url{https://xmm-tools.cosmos.esa.int/external/xmm_user_support/documentation/uhb/}}). Since extraction regions differ between observations, and the instrument response evolves over time, the effective area varies across observations. Thus, we calculated the ratio between the average effective areas of the three observations (Table~\ref{tab:areas and ratios}) and we applied them to the derived count rates (see Sect.~\ref{sec:energy-resolved light curves and hardness ratios} for more details).

\begin{table}
    \renewcommand{\arraystretch}{1.1}
    \caption{Ratios applied to the derived count rates to account for different effective areas in data extraction.}
    \centering
    \begin{tabular}{rrrrr} 
    \hline\hline
    \multicolumn{1}{c}
    {Obs} & 
    \multicolumn{1}{c}
    {1--3\,keV} 
    &
    \multicolumn{1}{c}
    {3--6\,keV} 
    &
    \multicolumn{1}{c}
    {6--8\,keV}
    &
    \multicolumn{1}{c}
    {8--10\,keV}
    \\
    \hline
    1 & 
    1.0000 & 1.0000 & 1.0000 & 1.0000 
    \\
    2 & 
    1.2738 & 1.1235 & 
    1.0969 & 1.0805 
    \\
    3 & 
    1.2740 & 1.1249 &
    1.0970 & 1.0784
    \\
    \hline
    \end{tabular}
    \label{tab:areas and ratios}
\end{table}

\subsection{Further software used}
\label{sec:further software used}

Besides the use of SAS to generate the event files, the data analysis and visualization of this paper were performed with Python v3.11.10 64-bit and, among others, the Astropy \citep{Astropycollab+2013} and Scipy \citep{Virtanen+2020} packages. We also used different functions from the spectral-timing software package Stingray\footnote{\url{https://docs.stingray.science/en/stable/}} \citep{Huppenkothen+2019,Bachetti+2024} v2.1\footnote{\url{https://zenodo.org/records/11383212}}, specifically \texttt{pulse.search.epoch\_folding\_search} to determine the pulse periods, \texttt{Lightcurve.make\_lightcurve} for light curve generation and \texttt{pulse.pulsar.fold\_events} for pulse profiles.

\begin{figure*}
    \centering
    \includegraphics[width=\textwidth]{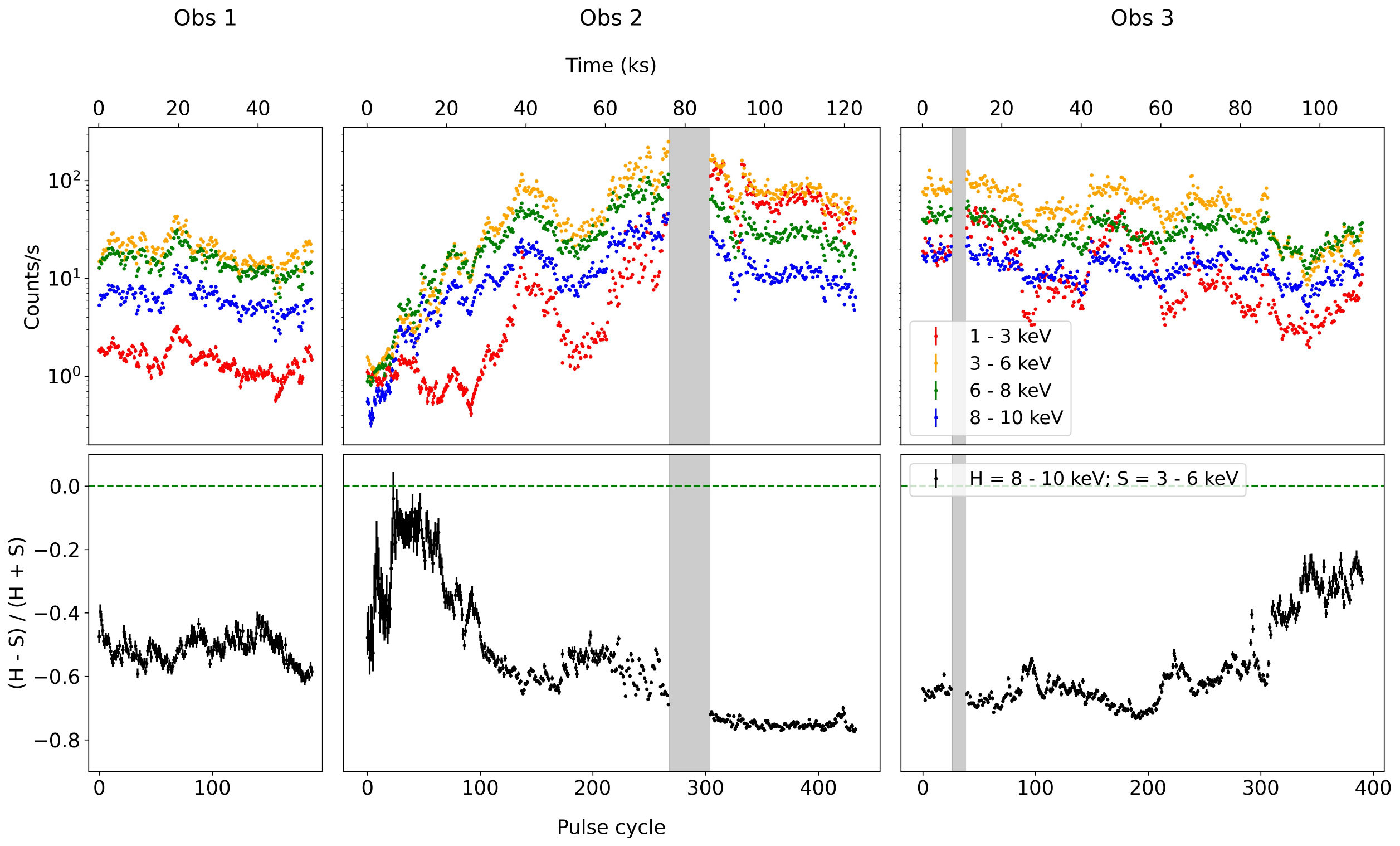}
    \caption{Energy-resolved light curves (top panels) and HRs (Eq.~\ref{eq:hardness ratio}) between the 8--10 and 3--6\,keV bands (bottom panels), with a time resolution of the respective pulse period, for the three observations. The width of each column is proportional to the duration of the corresponding observation. The light curves are plotted on a logarithmic scale for clarity. The colors represent different energy bands following the legend in the plot. Gray areas mark intervals of bright flares excluded for time-resolved analysis (see Sect.~\ref{sec:pulse periods and pulse cycles} and Table~\ref{tab:pulse periods, offsets and pulse ranges}).}
    \label{fig:energy-resolved light curves and hardness ratios}
\end{figure*}

\section{Light curves and pulse profiles per observation}
\label{sec:light curves and pulse profiles per observation}

\subsection{Pulse periods and pulse cycles}
\label{sec:pulse periods and pulse cycles}

For consistency, we redetermined pulse periods for all three observations, using 256 as the number of bins and as oversampling factor in the search. For Obs~2 and Obs~3, we used the published periods from \citet{Martinez+2014} and \citet{Diez+2023} as references, rounded to 283.39\,s and 283.45\,s, respectively. For Obs~1, we chose a reference value of 283.5\,s, interpolating between earlier measurements with \batse \citep[Fig.~8]{Kretschmar+2021} and the periods found by \citet{Staubert+2004} in \integral observations carried out in 2003. Uncertainties in the found periods were calculated using Eqs.~3 and~B2 in \citet{MartinCarrillo+2012}. The periods we found are shown in Table~\ref{tab:pulse periods, offsets and pulse ranges}. For what concerns Obs~2 and Obs~3, our values agree within the uncertainties with the values determined in the earlier publications. The pulse periods remain stable within uncertainties during the individual observations, since maxima and marked minima of individual pulse cycles align in phase throughout the whole time intervals covered.
 
For the study of pulse profiles and their behavior during individual pulse cycles, we explored different values of $2^n$ for the number of phase-bins and settled on 32 as a balance between resolution and typical significance of the count rate in each bin. In the following we mainly discuss data at the resolution of full pulse cycles. For coherence between different data sets, we used the first occurrence of a marked minimum in the profiles, visible in individual pulse trains, as effective zero point for each observation. These times are offset from the first arrival of each observation (51850.617, 53880.452, 58606.927\,MJD, for Obs~1, Obs~2 and Obs~3, respectively) as reported in Table~\ref{tab:pulse periods, offsets and pulse ranges}. Ignoring partial pulse cycles at the end of each observation, 189, 434 and 391 full pulse cycles are obtained for the three observations.

A further complication arose from bright flares in Obs~2 and Obs~3 saturating the onboard buffer and disturbing temporal information on timescales of individual pulse cycles \citep{Martinez+2014, Diez+2023}. For this reason, we excluded the data between $T=76$\,ks and $T=86$\,ks in Obs~2 and $T=7.5$\,ks and $T=11$\,ks in Obs~3. The final sets of pulse cycles used are listed in Table~\ref{tab:pulse periods, offsets and pulse ranges}. 

\begin{table}
    \renewcommand{\arraystretch}{1.1}
    \caption{Pulse periods, offset times from the first arrival and full pulse cycles used in the analysis. For Obs~2 and Obs~3 the ranges are before; after the flares excluded in the analysis.}
    \centering
    \begin{tabular}{rrrr@{--}l@{ ; }r@{--}l}    
    \hline\hline
    \multicolumn{1}{c}
    {Obs} & 
    \multicolumn{1}{c}{Pulse period (s)} 
    &
    \multicolumn{1}{c}{Offset (s)}
    &
    \multicolumn{4}{c}
    {Pulse cycle intervals}
    \\
    \hline
    1 & 
    $283.523 \pm 0.006$ & 
    148.30 & 
    \multicolumn{2}{c}{ } & 
     0 & 188 
    \\  
    2 & 
    $283.395 \pm 0.003$ &  33.51 &
    0 & 267 &
    304 & 433 
    \\
    3 & 
    $283.447 \pm 0.003$ & 236.87 & 
    0 & 25 &
    39 & 390 \\
    \hline          
    \end{tabular}
    \renewcommand{\arraystretch}{1.0}
    \label{tab:pulse periods, offsets and pulse ranges}
\end{table}

\begin{figure}
    \centering
    \includegraphics[width=0.5\textwidth]{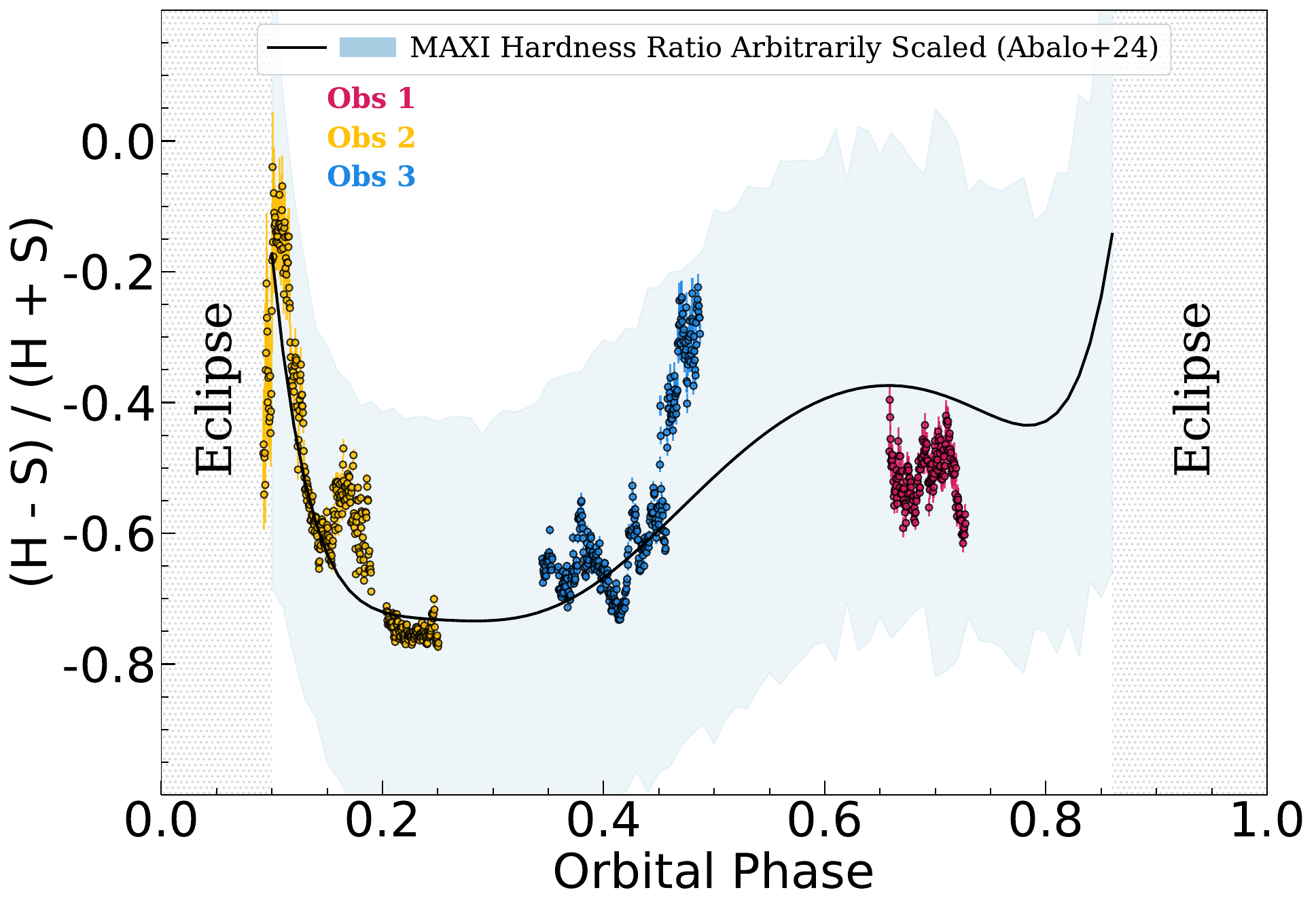}
    \caption{HR evolution during the three \xmmnewton observations of Vela~X-1 compared to the orbit-averaged \textit{MAXI}/GSC trend (2009–2024), adapted from \citet{Abalo+2024}. \textit{MAXI} HRs, based on Crab-like spectra, appear systematically harder and are rescaled for comparison with \xmmnewton, which uses count-based HRs.}
    \label{fig:hr_context}
\end{figure}

\begin{figure*}
    \centering
    \includegraphics[width=\textwidth]{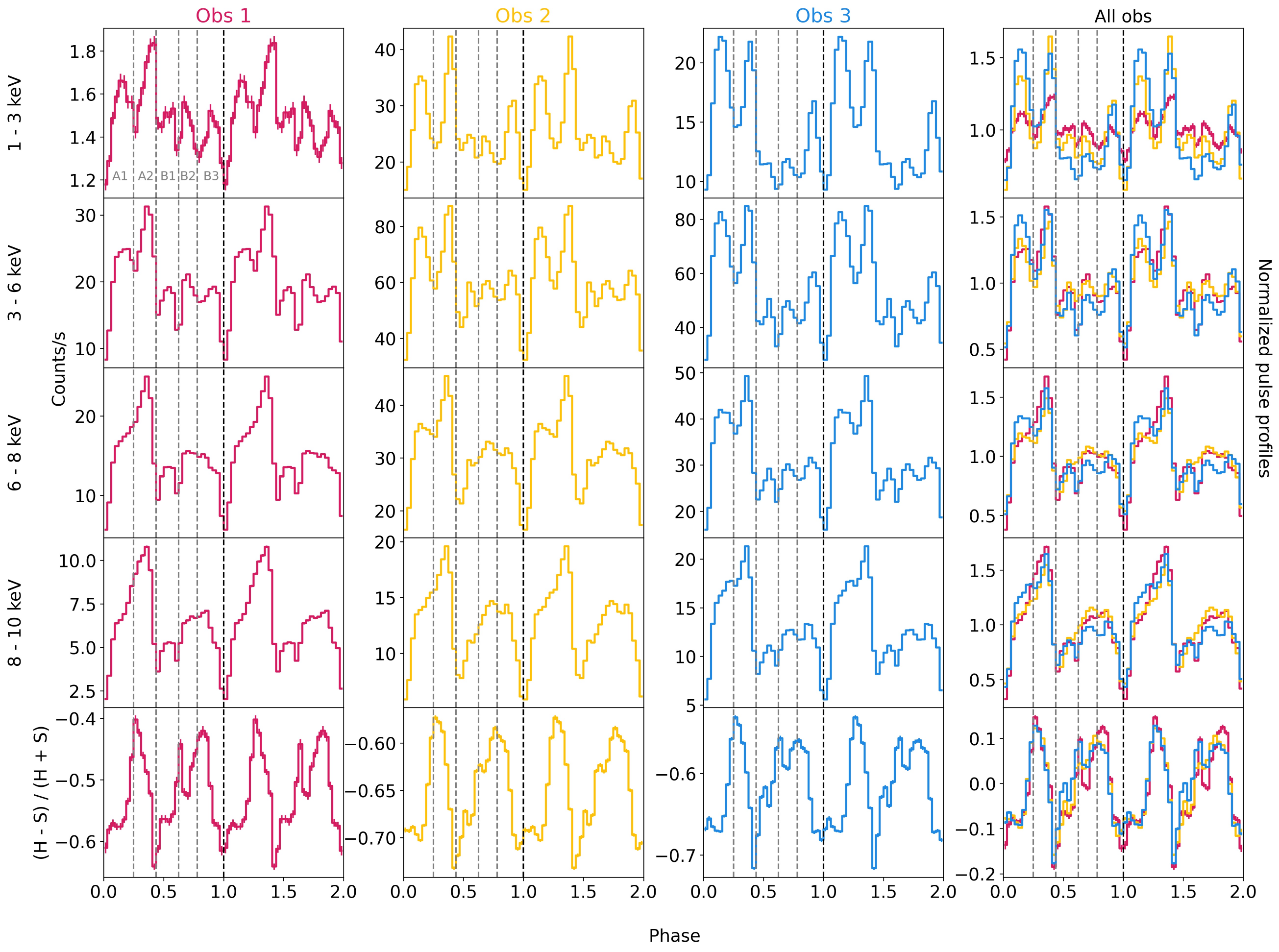}
    \caption{Mean pulse profiles of the three observations in the four energy bands used throughout this article and the corresponding HRs between the 8--10 and the 3--6\,keV bands. In the rightmost column, we compare the profiles across observations, normalizing the individual profiles by their respective mean count rates. The error bars are smaller than the line width in nearly all of the panels. Dashed lines indicate the five defined phase regions, $A1$ to $B3$, as explained in the text.}
    \label{fig:pulse profiles}
\end{figure*}

\subsection{Energy-resolved light curves and hardness ratios}
\label{sec:energy-resolved light curves and hardness ratios}

Following \citet{Martinez+2014}, we generated light curves in  four energy bands: 1--3, 3--6, 6--8 and 8--10\,keV. Poisson uncertainties were calculated using \texttt{stats.poisson\_conf\_interval}\footnote{\url{https://docs.astropy.org/en/stable/api/astropy.stats.poisson_conf_interval.html}} with ``frequentist-confidence'' as interval \citep[for further details, see][]{Maxwell+2011} and $\sigma = 1$. To correct for the differences in effective areas (Sect.~\ref{sec:data reduction}), we normalized the derived count rates by the factors listed in Table~\ref{tab:areas and ratios}. 

In order to track changes in the source spectrum as a function of time, we used the hardness ratio (HR), defined as: 
\begin{equation}
    \mathrm{HR} = \dfrac{H - S}{H + S} 
    \label{eq:hardness ratio}
\end{equation}
where $H$ and $S$ denote the flux in the hard and soft band, respectively. In the following, the HR is always calculated between the 8--10\,keV (hard) and the 3--6\,keV (soft) bands, since this HR has been found to closely follow the evolution of the photoelectric absorption $N_\mathrm{H}$ obtained from spectral fitting for Obs~2 \citep[Figs.~4 and~7]{Martinez+2014} and Obs~3 \citep[Figs.~5 and~9]{Diez+2023}. 

We present the energy-resolved light curves and the HR curves for the three observations in Fig.~\ref{fig:energy-resolved light curves and hardness ratios}. To contextualize the HR evolution observed in our \xmmnewton data, Fig.~\ref{fig:hr_context} compares it to the long-term, orbit-averaged trend derived from \textit{MAXI}/GSC observations \citep{Abalo+2024}. A detailed discussion of this comparison is provided in Sect.~\ref{sec:strong fluctuations below 3 keV: variable absorption in large structures}.

Obs~1 shows relatively mild variability by a factor of about 6 in flux. The HR is around a mean value of about $-$0.5, and shows limited variability with a slight decay at the end of the observation.

Obs~2 is much more variable. The first data points are from a time when the neutron star was coming out of eclipse, a time period partially ignored in \citet{Martinez+2014} to avoid times close to the radiation belts, but included here since we concentrate on timing instead of spectral analysis. Even outside the excluded flaring period, observed fluxes vary by a factor of more than 360 in the 1--3\,keV energy band or $\sim$140 in the 8--10\,keV range. The HR shows a strong evolution, with both the highest and the lowest values across all data sets reached in this observation. The impact of the correlated changes in absorption are clearly visible in the flux evolution of the light curve at 1--3\,keV.

Obs~3 presents an intermediate picture between the two other observations. Flux varies by up to a factor of $\sim$27 in the 1--3 keV range and $\sim$6 in the 8--10 keV band. The HR rises markedly in the second half of the observation, which has been associated to the signature of the accretion wake moving across the line of sight \citep{Diez+2023}.

\subsection{Energy-resolved pulse profiles and hardness ratios}
\label{sec:energy-resolved pulse profiles and hardness ratios}

We present in Fig.~\ref{fig:pulse profiles} the energy-resolved pulse profiles averaged over the time intervals taken into account for each observation (Table~\ref{tab:pulse periods, offsets and pulse ranges}), along with their corresponding HRs. Since the profiles were obtained from the events and not from the light curves, we applied again the factors from Table~\ref{tab:areas and ratios}. A comparison between the normalized pulse profiles of the three observations is displayed in the last column, where the normalization was performed by dividing the individual profiles by their mean rate. We applied this method to all the normalized pulse profiles shown in this article.

For the further discussion we defined two parts of the profile, $A$ and $B$, associated with the two peaks visible at higher energies, which are generally linked to emission from the two polar regions. Based on the 32 bins used throughout this study, the boundary between these parts is right after the $14^{\textrm{th}}$ phase-bin. Motivated by the sub-peaks, visible especially below 3\,keV, we further split these intervals, such that $A$ is formed by the two sections $A1$ (bins 0--7) and $A2$ (8--13), while $B$ consists of sections $B1$ (14--19), $B2$ (20--24) and $B3$ (25--31). These boundaries are indicated in Fig.~\ref{fig:pulse profiles}. This segmentation is similar to the one used by \citet{Raubenheimer+1990}, but with flipped definition of $A$ and $B$, and their region $C$ roughly corresponding to our $B1$.

While generally similar, the profiles shown in Fig.~\ref{fig:pulse profiles} display differences at the level of these components. $A2$ almost always contains the maximum of the profiles, except for the 1--3\,keV profile of Obs~3 and, in general, the ratio between $A1$ and $A2$ varies between different observations. $B1$, $B2$ and $B3$ also show variations in their relative strength as function of energy and between observations.

The last column of Fig.~\ref{fig:pulse profiles} confirms the general stability of the pulse profile shape across many years and different levels of source brightness. The main differences are at low energies, where pulse amplitudes are reduced, especially for Obs~1. 

The bottom panels of Fig.~\ref{fig:pulse profiles} show the HRs between the 8--10 and 3--6\,keV profiles. For all observations, the HR minimum coincides in pulsed phase with the boundary between $A$ and $B$, while the maxima coincide with the boundaries between the subpeaks. 

\begin{figure*}
    \centering
    \includegraphics[width=\textwidth]{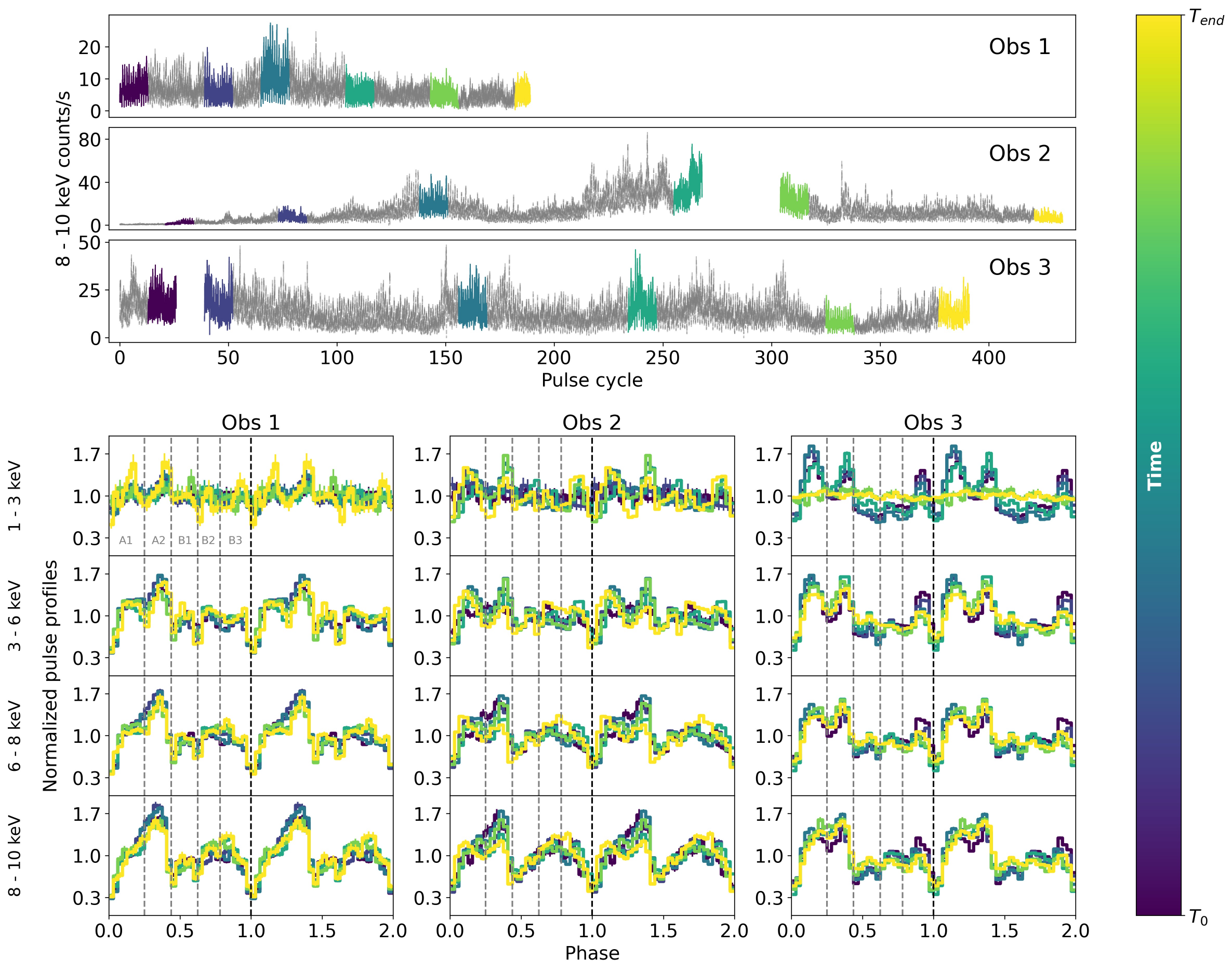}
    \caption{Top panels: 8--10\,keV light curves with a time resolution of an individual phase-bin ($\sim$ 8.86~s). We mark in color the time ranges used to build the time-resolved profiles shown in the lower panels.}
    \label{fig:time-resolved plots}
\end{figure*}

\section{Pulse profile variations with time and flux}
\label{sec:pulse profile variations with time and flux}

The mean pulse profiles in Fig.~\ref{fig:pulse profiles} average data over extended periods with substantial flux and hardness variability (Fig.~\ref{fig:energy-resolved light curves and hardness ratios}). If the profile shape depends on the flux, mean shapes will be dominated by the brightest pulse cycles. We next examine how pulse shape evolves with time and varies with flux.

\subsection{Variations with time within individual observations}
\label{sec:variations with time within individual observations}

We studied the evolution of the energy-resolved pulse profiles within each observation. For this analysis we aimed to split the total number of pulse cycles in each observation (Table~\ref{tab:pulse periods, offsets and pulse ranges}) as evenly as possible, while finding a compromise between the goals of higher time resolution to follow variations in the light curves and statistical quality requiring integration over a sufficient number of pulse periods. We settled on a baseline of 13 pulse periods per profile, corresponding to $\sim 1$\,h of integration time, with some variations as explained below. This choice was also driven by the will to have profiles of equal integration time before and after the bright flares excluded in the analysis (Sect.~\ref{sec:pulse periods and pulse cycles}). All of the above led to the following splits per observation. For Obs~1, there are 14 full-length profiles and a final one covering 7 pulse periods. Regarding Obs~2, up to the flare, we have one profile covering 8 pulse periods and 20 full-length profiles; after the flare, 10 full-length profiles. For Obs~3, there are 2 full-length profiles before the flare; after this, 26 full-length profiles and one covering 14 pulse periods. In Fig.~\ref{fig:time-resolved plots}, we can visualize the changes in profile shapes during the observations, with example profiles at different times marked in the light curves shown above.

The most pronounced pulse profile changes occur in the 1--3\,keV band, where pulsations are nearly absent for most of Obs~1 --- except near the end when HR drops (Fig.~\ref{fig:energy-resolved light curves and hardness ratios}) --- and similarly in the initial and final time ranges of Obs~2 and Obs~3, respectively, when HR is also high. Profiles at higher energies remain more stable but still show variations in the relative brightness between parts ($A$ and $B$) and between sections ($A1$, $A2$, $B1$, $B2$, $B3$). For Obs~1, the amplitude of $A$ slightly decreases with respect to that of $B$ as time goes on. This is even more pronounced in Obs~2, where the amplitude of $B$ surpasses that of $A$ at the end of the observation. For Obs~3, $A1$ and $B3$ are relatively brighter in the earlier profiles than in the later ones.

\begin{figure}
    \centering
    \includegraphics[width=0.5\textwidth]{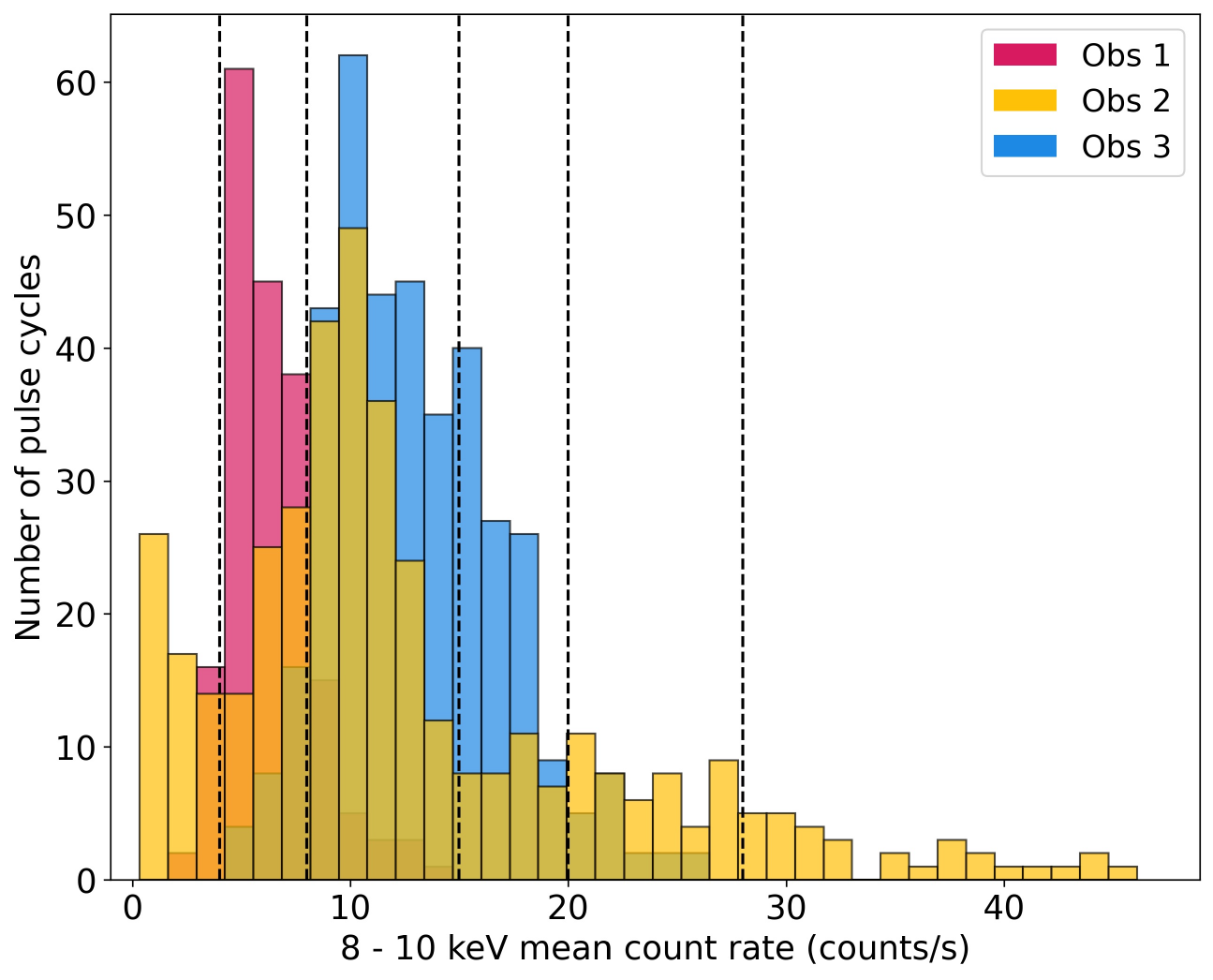}
    \caption{Distributions of the 8--10\,keV mean count rate per pulse cycle. 
    The black dashed lines indicate the boundaries between flux intervals.
    The pulse profiles obtained for each flux interval and observation are shown in Fig.~\ref{fig:flux-resolved plots}.}
    \label{fig:mean distribution}
\end{figure}

\begin{figure*}
    \centering
    \includegraphics[width=\textwidth]{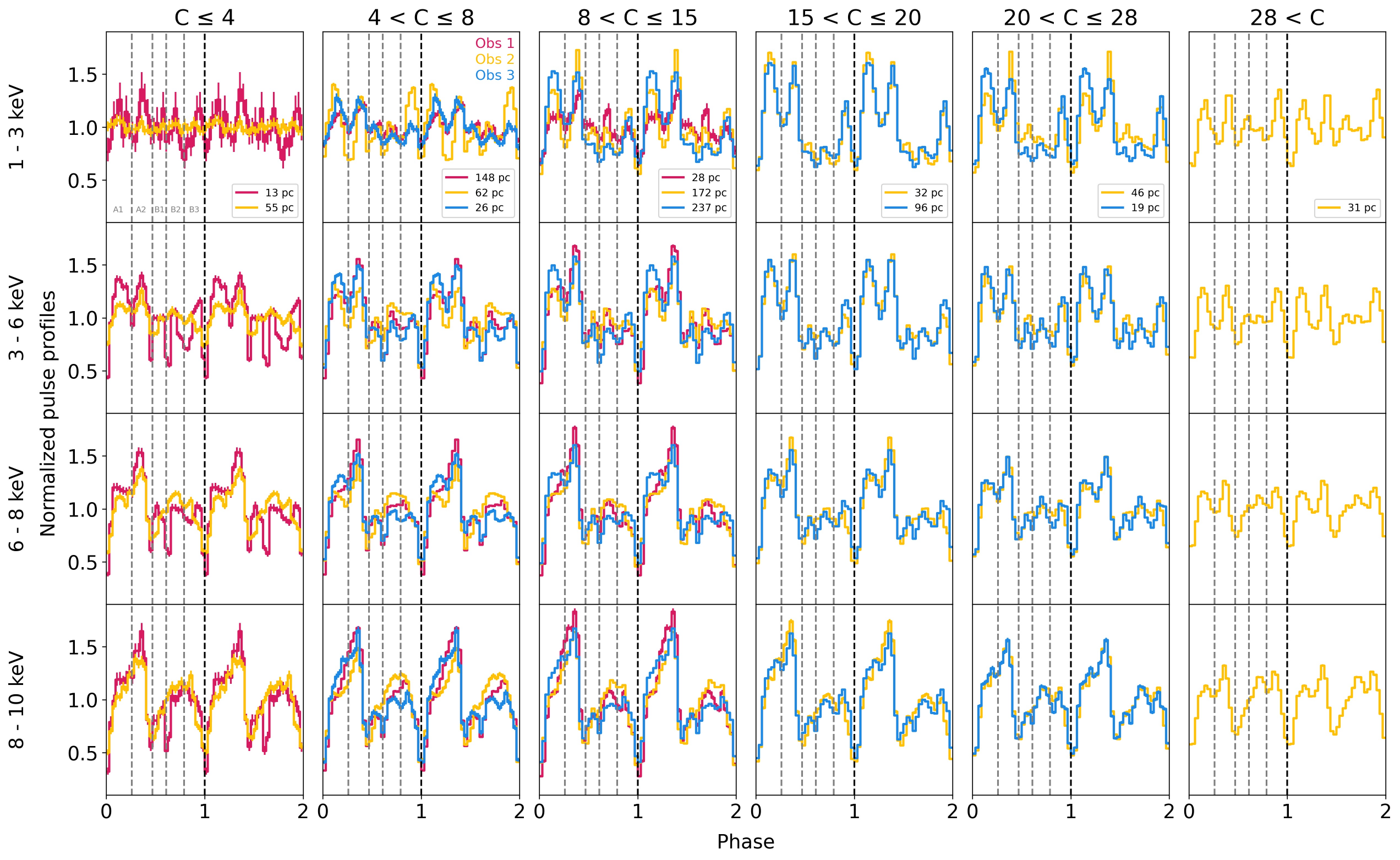}
    \caption{Normalized flux-resolved pulse profiles obtained as explained in Sect.~\ref{sec:variations with flux within and across observations}. Each column corresponds to a different flux range, given by the count rate $C$ (counts/s) in the 8--10\,keV band averaged over the pulse cycle. The number followed by ``pc'' in the top row is the number of pulse cycles per observation and flux level. Obs~2 covers the whole flux range, while Obs~1 and Obs~3 only have data within 3 and 4 flux intervals, respectively.
    The distributions of the 8--10\,keV mean count rate per pulse cycle and the flux limits are shown in Fig.~\ref{fig:mean distribution}.}
    \label{fig:flux-resolved plots}
\end{figure*}

\subsection{Variations with flux within and across observations}
\label{sec:variations with flux within and across observations}

To investigate a possible relation between the source flux and the observed pulse profile shape, we classified all pulse cycles by their mean count rate in the 8--10\,keV energy range. We selected this energy band as it is the least affected by absorption. We chose count rate intervals covering the observed range variability and ensuring that each observation was covered by at least three intervals. This led to the following count rate boundaries between intervals: 4, 8, 15, 20, and 28\,counts/s (Fig.~\ref{fig:mean distribution}).

The flux-resolved pulse profiles for different observations (Fig.~\ref{fig:flux-resolved plots}) are visibly more similar for higher fluxes. The similarity across observations also appears to increase with energy. Note that the pulse profiles averaged over the brightest pulse cycles -- a range only covered within Obs~2 -- show a significantly different shape, with $A$ and $B$ being almost equally bright.

\begin{figure}
    \centering
    \includegraphics[width=0.5\textwidth]{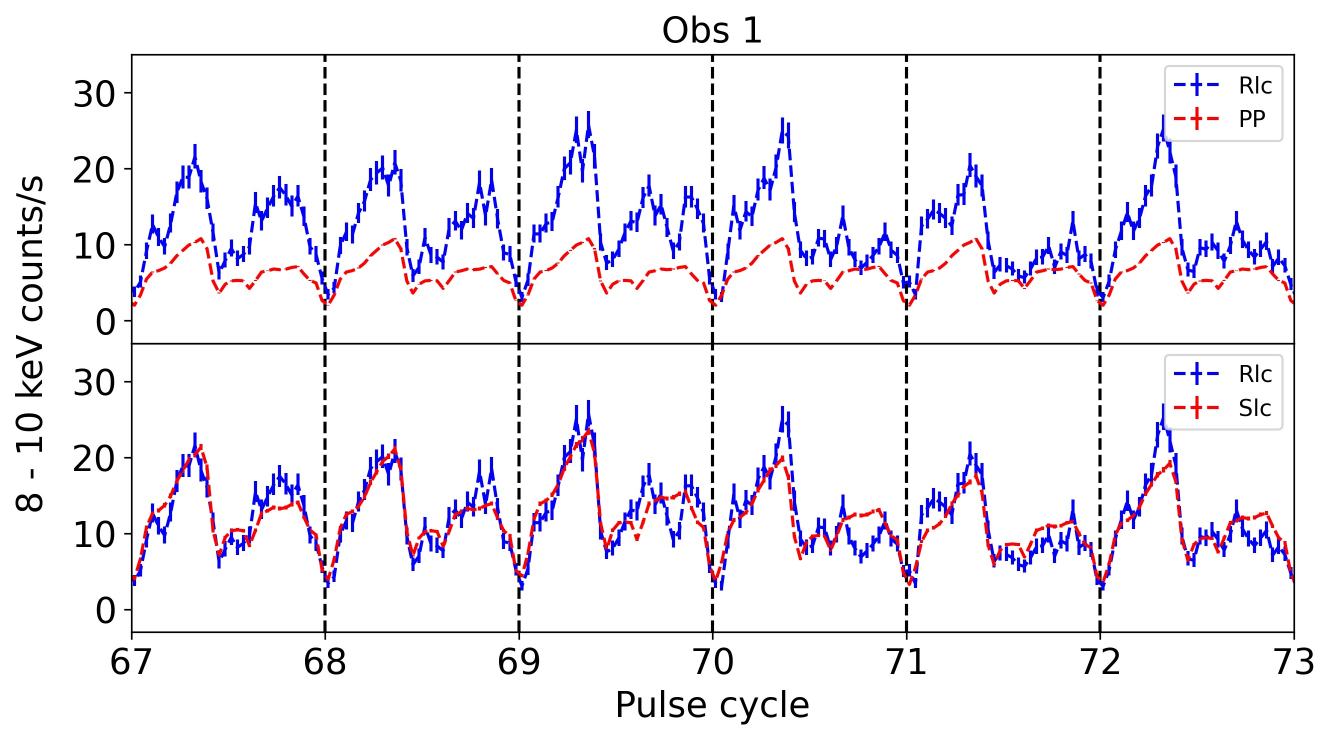}
    \caption{Demonstration of the approach to generate the Slc by scaling the repeated PP to match the mean flux of the corresponding pulse cycle of the Rlc. The top panel shows the Rlc and the repeated PP before scaling, while the bottom panel shows the Rlc and the final Slc. See Sect.~\ref{sec:synthetic light curves scaling the mean profile} for more details.}
    \label{fig:scaling method}
\end{figure}

\section{Variations of individual pulse cycles}
\label{sec:variations of individual pulse cycles}

At least some of the variations between pulse profiles described in the previous sections will be driven by the known variations between pulse cycles, which are also visible in the light curves of all three observations. In the following, we quantify these variations.

\subsection{Total flux variations between consecutive pulse cycles}
\label{sec:total flux variations between consecutive pulse cycles}

As a measure for intrinsic variations in X-ray brightness from one pulse cycle to the next, we calculated the ratio of total flux in the 8--10\,keV band between consecutive pulse cycles. For Obs~1, the minimum and maximum ratios are 0.61 and 1.75, respectively, while 90\% of the values lie in the range 0.78--1.26. For Obs~2, the corresponding ratios are 0.58 and 2.25 and 0.75--1.37, while for Obs~3, they are 0.60 and 1.58 and 0.79--1.28. 

\subsection{Variable contributions from the individual pulse sections}
\label{sec:variable contributions from the individual pulse sections}

The fraction of phase covered by $A$ and $B$ is 0.4375 and 0.5625, respectively. However, the fraction of flux is almost perfectly identical on average. Considering all individual pulse cycles of all observations and energy bands, the mean flux fraction in $A$ is 0.498, while the median is 0.499 and for 90\% of the pulse cycles this fraction is in the range 0.408--0.587, i.e., a scatter of about 10\% from the even distribution.

We also calculated the flux fraction per pulsed phase section ($A1$, $A2$, $B1$, $B2$ and $B3$) in each pulse cycle. The mean and standard deviation of these fractions confirm that the contribution of each section to the total emission is fairly constant across all observations and energy bands, with variations of the order of 10--20\% of the average. $A1$ and $B1$ tend to diminish with energy, while $A2$ and $B2$ tend to rise. $B3$ does not seem to evolve with energy range.

Overall variations may not be representative of fluctuations between consecutive pulse cycles. Therefore, we quantified the variability of each pulsed phase section ($X$) from one pulse cycle to the next with:
\begin{equation}
    \mathrm{R}_{X,(j+1)/j} = \dfrac{({F}_{X} / {F}_{T})_{j+1}}{({F}_{X} /{F}_{T})_{j}}
    \label{eq:ratio flux per section}
\end{equation}
where $j$ is the pulse count, $F_{X}$ is the flux in the selected section and $F_{T}$ is the total flux in the selected pulse cycle. Considering all energy bands, $A2$ fluctuates less from one pulse cycle to the next than the rest of the pulsed phase sections, with the geometric standard deviation\footnote{\url{https://scipy.github.io/devdocs/reference/generated/scipy.stats.gstd.html}} of $\mathrm{R}_{A2}$ between 1.14 and 1.25. On the opposite extreme we find $B2$ and $B3$, with the corresponding geometric standard deviations in the ranges 1.17--1.26 and 1.19--1.32, respectively.  

\subsection{Variations in shape compared to the mean profile}
\label{sec:variations in shape compared to the mean profile}

\subsubsection{Synthetic light curves scaling the mean profile}
\label{sec:synthetic light curves scaling the mean profile}

\citet{Kretschmar+2014} compared the extracted light curves with ``ideal'' light curves, which were obtained in two steps. They first repeated the mean pulse profile as many times as pulse cycles were in the extracted light curve. The second step was to scale the predicted curve to the extracted one, so that they were comparable regardless of the flux variability. 

Following a similar approach, we created ``synthetic light curves'' (Slc) by repeating the mean pulse profile (PP) for each observation sufficient times to cover the pulse cycles used (Table~\ref{tab:pulse periods, offsets and pulse ranges}), normalizing this curve and multiplying by the mean count rate of the real light curve (Rlc) for each pulse cycle:
\begin{equation}
    \mathrm{Slc}_{i,j} = \dfrac{ \mathrm{PP}_{i} }{\overline{\mathrm{PP}_{i}}}\times \left(\overline{\mathrm{Rlc}_{i}}\right)_{j}
    \label{eq:scaling method}
\end{equation}
where $i$ denotes the phase-bin. 

We also tried other scaling approaches, including separate scale factors for the regions $A$ and $B$, but we finally chose this one as it produced the fewest artifacts when visually inspected. An example comparing a few pulse cycles of real data with a simple repeated profile and the scaled version is shown in Fig.~\ref{fig:scaling method}.

\subsubsection{Quantitative comparison parameters}
\label{sec:quantitative comparison parameters}

To measure the degree of similarity between the Rlc and the Slc, we calculated the Pearson Correlation Coefficient (PCC) with the Scipy function \texttt{stats.pearsonr}\footnote{\url{https://docs.scipy.org/doc/scipy-1.16.0/reference/generated/scipy.stats.pearsonr.html}} (setting 0.9 as the confidence level). Using their definition of PCC, our coefficient per pulse cycle is:
\begin{equation}
    \mathrm{PCC}_{j} = \dfrac{\sum_{i=1}^{\textrm{N}} (\textrm{Rlc}_{i,j}- (\overline{\mathrm{Rlc}_{i}})_{j})\times{(\textrm{Slc}_{i,j}-(\overline{\mathrm{Slc}_{i}})_{j}})}
    {\sqrt{{\sum_{i=1}^{\textrm{N}}( \textrm{Rlc}_{i,j}-(\overline{\mathrm{Rlc}_{i}})_{j})^{2}\times{\sum_{i=1}^{\textrm{N}}( \textrm{Slc}_{i,j}- (\overline{\mathrm{Slc}_{i}})_{j})^{2}}}}}
    \label{eq:pcc}
\end{equation}
where N is the number of bins per pulse cycle. PCC can take any value between -1 (perfect negative correlation) and 1 (perfect positive correlation). PCC = 0 means that there is no correlation at all.

To quantify the resemblance between the curves on timescales of the duration of an individual bin, we measured the absolute percentage error (APE):
\begin{equation}
    \textrm{APE}_{i,j} = \dfrac{\left| \textrm{Rlc}_{i,j}- \textrm{Slc}_{i,j}\right|}{\textrm{Rlc}_{i,j}}
    \label{eq:ape}
\end{equation}

\subsubsection{Distribution of PCC values per pulse cycle}
\label{sec:distribution of PCC values per pulse cycle}

\begin{figure*}
    \centering
    \includegraphics[width=\textwidth]{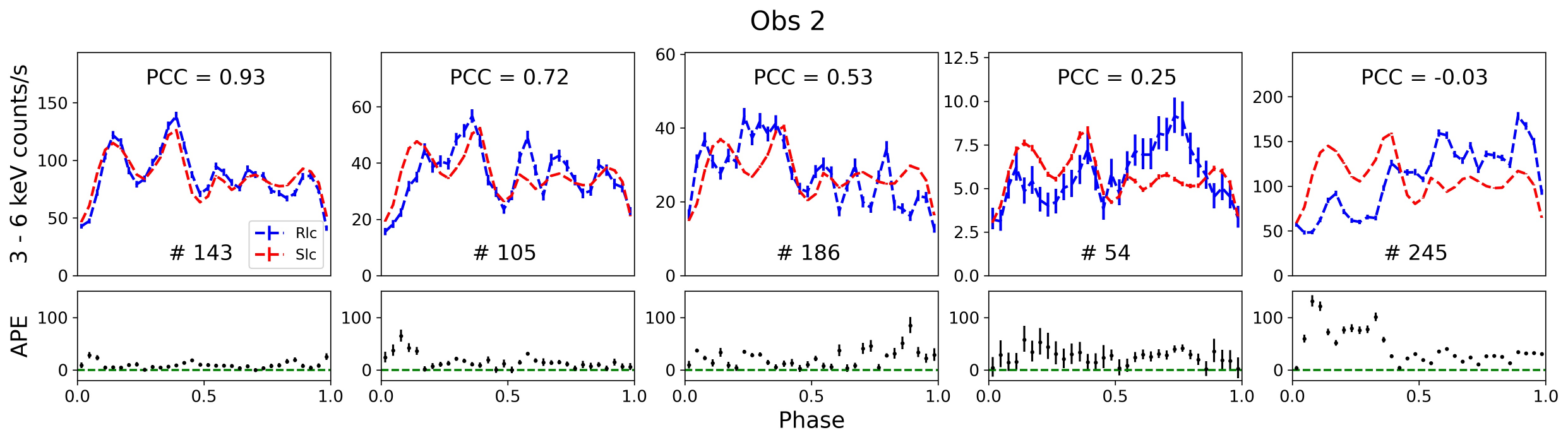}
    \caption{Comparison between the Rlc and the Slc for some pulse cycles of Obs~2 in the 3--6\,keV range. The pulse cycle number is indicated right after the symbol \# in the top panels. The generation of the Slc is explained in Sect.~\ref{sec:synthetic light curves scaling the mean profile}. The similarity between the Rlc and the Slc decreases from left to right, as quantified by PCC (obtained with Eq.~\ref{eq:pcc}) and APE (obtained with Eq.~\ref{eq:ape}). }
    \label{fig:rlc vs slc examples}
\end{figure*}

The PCC values for individual pulse cycles range from somewhat anticorrelated ($-$0.47) to almost perfectly matching (0.97), showing a wide variety of shapes for individual pulse cycles. Examples of good and bad matches between the curves are shown in Fig.~\ref{fig:rlc vs slc examples}.

\begin{figure*}
    \centering
    \includegraphics[width=\textwidth]{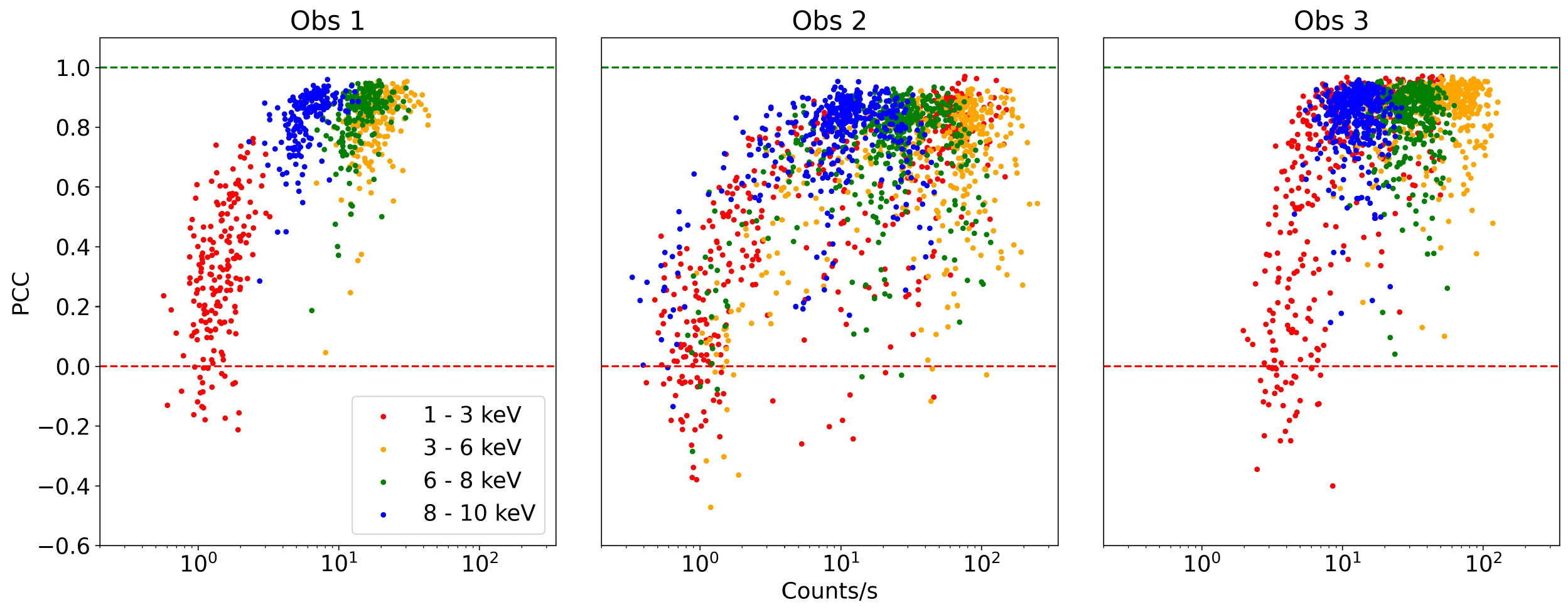}
    \caption{PCC (obtained with Eq.~\ref{eq:pcc}) between the Rlc and the Slc per pulse cycle, plotted against the mean count rate (in logarithmic scale) of the corresponding pulse cycle. The colors represent different energy bands following the legend in the plot. Error bars are not shown for a clearer visualization of the scatter of individual points. The green and red lines mark the interval between perfect correlation and no correlation.}
    \label{fig:pcc vs mflux}
\end{figure*}

To assess the relation between these variations and flux, we plotted the PCC against the mean count rate (Fig.~\ref{fig:pcc vs mflux}). 

The Rlc and the Slc behave, in general, similarly, regardless of the observation. Nonetheless, Fig.~\ref{fig:pcc vs mflux} shows that the differences between the curves increase as flux decreases. Due to absorption, the lowest fluxes and highest differences are detected in the 1--3 keV range.

To further quantify the difference in shape between the Rlc and Slc pulse cycles, we calculated the 0.1 and 0.9 quantiles ($Q_{\textrm{0.1}}$ and $Q_{\textrm{0.9}}$, respectively) of the PCC distributions (Table~\ref{tab:pcc quantiles}). In this analysis, we sought to exclude pulse cycles without evident pulsations. At the beginning of Obs~2, with the neutron star coming out of the eclipse, the signal of individual pulse cycles is not significantly different from a constant close to zero, even at the highest energies. Making a cut at a mean count rate of 2~counts/s in the 8--10\,keV band, we excluded pulse cycles 0--26, 41 and 44 of Obs~2 (eclipse egress started between pulse cycles 21--49), while the other observations were not affected.

\begin{table}
    \renewcommand{\arraystretch}{1.1}
    \setlength{\tabcolsep}{5pt}
    \caption{0.1 and 0.9 quantiles of the PCC distributions, ignoring pulse cycles with mean count rate lower than 2~counts/s in the 8--10\,keV range.}
    \centering
    \begin{tabular}
    {r|rr|rr|rr|rr} 
    \hline\hline
    \multirow{2}{*}{Obs} 
    &
    \multicolumn{2}{c|}
    {1--3\,keV}
    & 
    \multicolumn{2}{c|}
    {3--6\,keV}
    &
    \multicolumn{2}{c|}
    {6--8\,keV}
    &
    \multicolumn{2}{c}
    {8--10\,keV}
    \\
    &
    \multicolumn{1}{c}
    {$Q_{\textrm{0.1}}$} 
    & 
    \multicolumn{1}{c|}
    {$Q_{\textrm{0.9}}$} 
    &
    \multicolumn{1}{c}
    {$Q_{\textrm{0.1}}$} 
    & 
    \multicolumn{1}{c|}
    {$Q_{\textrm{0.9}}$} 
    &
    \multicolumn{1}{c}
    {$Q_{\textrm{0.1}}$} 
    & 
    \multicolumn{1}{c|}
    {$Q_{\textrm{0.9}}$} 
    &
    \multicolumn{1}{c}
    {$Q_{\textrm{0.1}}$} 
    & 
    \multicolumn{1}{c}
    {$Q_{\textrm{0.9}}$} 
    \\
    \hline
    1 & 
    -0.01 & 0.60 &
    0.68 & 0.92 &
    0.71 & 0.93 &
    0.70 & 0.92
    \\
    2 &
    0.05 & 0.88 &
    0.42 & 0.88 &
    0.54 & 0.89 & 
    0.62 & 0.90
    \\
    3 & 
    0.05 & 0.93 &
    0.70 & 0.94 &
    0.70 & 0.93 & 
    0.72 & 0.92
    \\
    \hline
    \end{tabular}
    \label{tab:pcc quantiles}
\end{table}

The lowest $Q_{\textrm{0.1}}$ values for all observations are in the 1--3\,keV range. The highest $Q_{\textrm{0.1}}$ values in Obs~2 and Obs~3 correspond to the 8--10\,keV range, while the highest $Q_{\textrm{0.9}}$ values are distributed between the 6--8\,keV (Obs~1), 8--10\,keV (Obs~2) and 3--6\,keV (Obs~3) bands. In summary, the Rlc and the Slc are least similar in the 1--3\,keV and most similar in the 8--10\,keV band. Still, the 3--6 and 6--8\,keV values are very close to those of the highest energy band. Comparing the observations, Obs 3 has overall the quantile values closest to 1, while Obs~2 is the furthest away.

\begin{figure}
    \centering
    \includegraphics[width=0.5\textwidth]{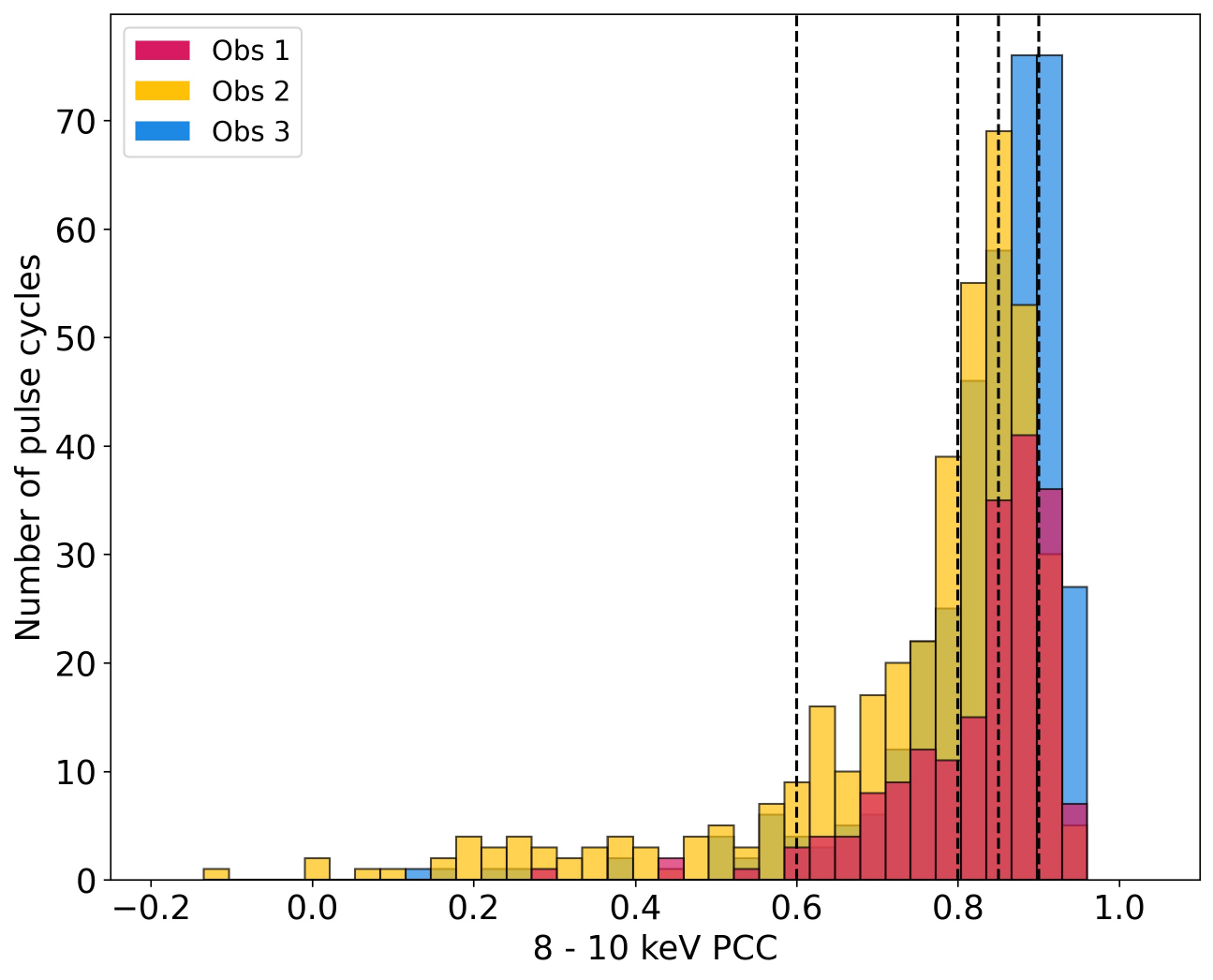}
    \caption{Distributions of the 8--10\,keV PCC (Eq.~\ref{eq:pcc}) between the Rlc and the Slc. The black dashed lines indicate the limits between PCC ranges. The PCC-resolved light curves and pulse profiles are shown in Fig.~\ref{fig:pcc-resolved plots}.}
    \label{fig:8-10 keV pcc distribution}
\end{figure}

\subsubsection{PCC-resolved pulse profiles}
\label{sec:pcc-resolved pulse profiles}

\begin{figure*}
    \centering
    \includegraphics[width=\textwidth]{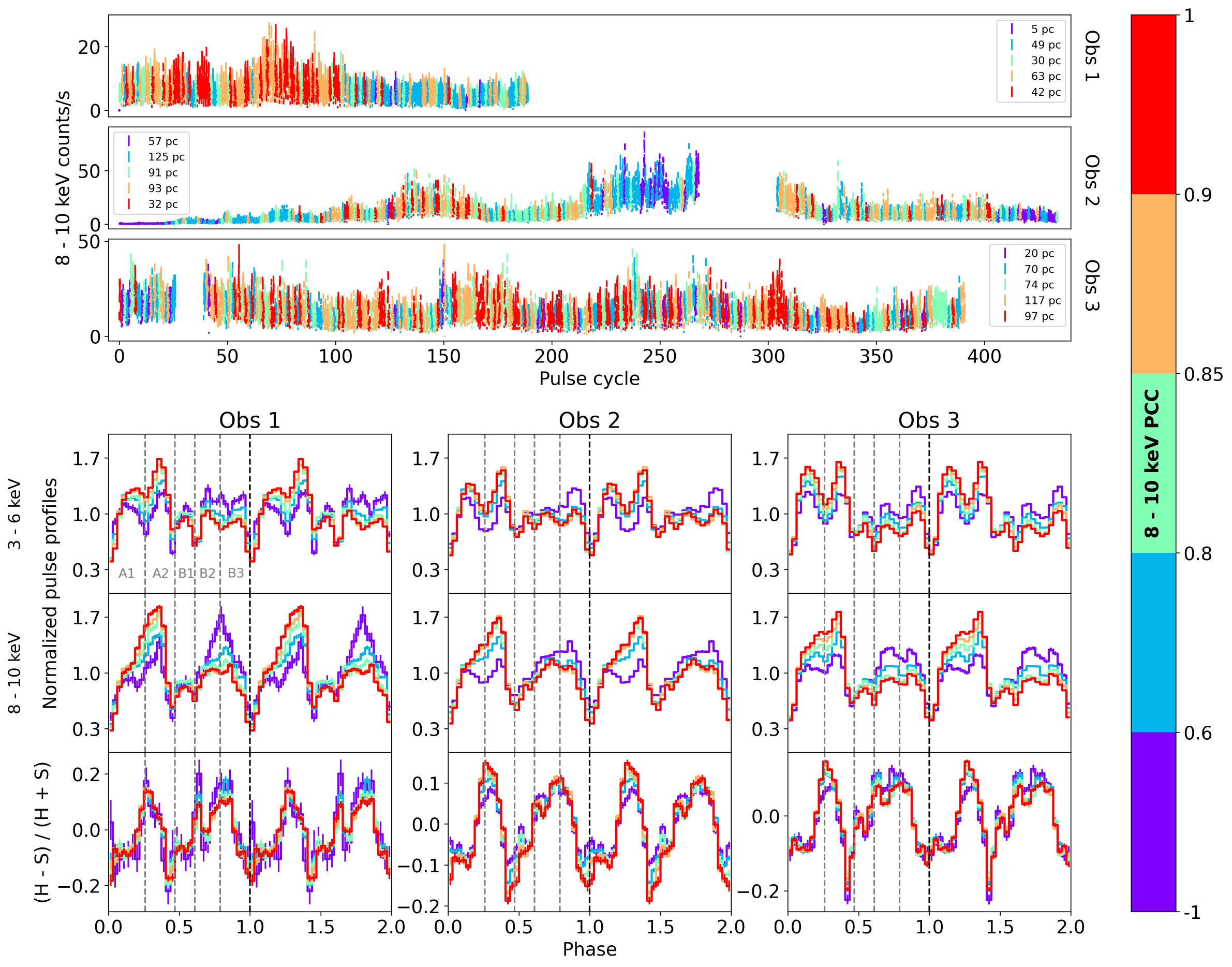}
    \caption{Top panels: 8--10\,keV PCC-resolved (see Sect.~\ref{sec:pcc-resolved pulse profiles}) light curves with a time resolution of an individual phase-bin ($\sim$ 8.86~s). The numbers followed by ``pc'' are the number of pulse cycles per PCC$_{8-10}$ range. First two rows of the bottom panels: normalized PCC-resolved pulse profiles in the 3--6 and 8--10\,keV bands. Last row of the bottom panel: HR between the 8--10 and 3--6\,keV normalized pulse profiles.}
    \label{fig:pcc-resolved plots}
\end{figure*}

In order to test the connection between PCC values and systematic shifts in the emission pattern, we chose to create ``PCC-resolved pulse profiles''. To do so, we divided the range in PCC values found in the 8--10\,keV band (labeled PCC$_{8-10}$ in the following) across all three observations into five intervals with PCC$_{8-10}$ values of 0.6, 0.8, 0.85 and 0.9 as limits (Fig.~\ref{fig:8-10 keV pcc distribution}). This choice ensured to have at least some pulse cycles of each observation within the chosen intervals. We then created mean profiles from the individual pulse cycles falling within a certain PCC$_{8-10}$ range, as shown in Fig.~\ref{fig:pcc-resolved plots}.

The light curves in Fig.~\ref{fig:pcc-resolved plots} show no evident temporal pattern for PCC$_{8-10}$ in any observation. The highest PCC$_{8-10}$ values usually correspond to pulse cycles of high flux; except for pulse cycles $\sim$ 230--267 of Obs~2, where during a very active period before the omitted massive flare, individual pulse cycles also deviate strongly from the mean shape. Regarding the PCC-resolved profiles, the amplitude of $A1$ and $A2$ increase with respect to that of $B2$ and $B3$ (in both energy bands) as PCC$_{8-10}$ increases, while the amplitude of $B1$ is much more stable. Note that the HR shape barely varies with PCC$_{8-10}$. It is also worth mentioning that, in the 8--10 keV range, the PCC-resolved profile from the  lowest PCC$_{8-10}$ range of Obs~1 is significantly different to that of Obs~2 and Obs~3. We visually confirmed that this is due to the small number of pulse cycles (5) combined for Obs~1 in this range.

\section{Discussion}
\label{sec:discussion}

\subsection{Overall pulse profile stability}  
\label{sec:overall pulse profile stability}

The analysis of three observations of Vela~X-1 with the same X-ray telescope, \xmmnewton, widely spaced in time (2000, 2006 and 2019) and at different orbital phases and luminosities, confirms again the long-term stability of the general pulse profile shape, and thus emission properties, averaged over many pulse cycles. Smaller, but significant differences remain in the strength of the visually defined subcomponents of the profiles (Fig.~\ref{fig:pulse profiles} and  Sect.~\ref{sec:variable contributions from the individual pulse sections}).

\subsection{Impact of absorption}
\label{sec:strong fluctuations below 3 keV: variable absorption in large structures}

Figure~\ref{fig:hr_context} places the HR evolution from our three \xmmnewton observations within the broader orbital context of Vela~X-1, as illustrated by the long-term averaged \textit{MAXI} curve from \citet{Abalo+2024}. The \textit{MAXI} trend shows systematic modulation across the orbit, with pronounced hardening near eclipse ingress and egress due to increased absorption in dense structures. The \xmmnewton tracks align with this behavior: Obs~1 and the end of Obs~3 coincide with phases of high obscuration, likely caused by wake structures, while the start of Obs~2 shows elevated HRs as the neutron star emerges from eclipse behind the extended stellar atmosphere.

Significant variations are observed in the ``soft'' pulse profiles within individual observations (Fig.~\ref{fig:time-resolved plots}), driven by changes in HR and thus $N_{\textrm{H}}$ (Figs.~\ref{fig:energy-resolved light curves and hardness ratios} and~\ref{fig:hr_context}). Increased absorption and scattering at lower energies lead to large count rate fluctuations and, at times, to a complete loss of the typical pulse shape. This behavior is reflected in the high scatter of individual pulse cycles around the mean profile, particularly in the 1–3\,keV band, as quantified by the low PCC values (Fig.~\ref{fig:pcc vs mflux}, Table~\ref{tab:pcc quantiles}).

The almost flat 1--3\,keV pulse profiles in Fig.~\ref{fig:time-resolved plots} are also a consequence of high absorption episodes. In Obs~1, the 1--3\,keV flux remains persistently low, consistent with dense wake structures crossing the line of sight at this orbital phase \citep{Kaper+1994, Doroshenko:2013, Abalo+2024}. As a result, pulsations are poorly defined in this energy range almost throughout the entire observation. A similar suppression is seen at the end of Obs~3, where comparable structures again obscure the neutron star \citep{Malacaria:2016, Diez+2023}. At the beginning of Obs~2, with the neutron star emerging from eclipse, photons were absorbed or scattered by the extended atmosphere of the supergiant \citep{Martinez+2014}. In all three cases, the duration of the suppressed pulsations suggests that large-scale, orbitally modulated structures dominate the absorption, rather than short-lived wind clumps.

\subsection{Flux dependence}
\label{sec:flux dependence}

Flux variations can yield changes in the height of the accretion column, leading to changes in the fraction of photons affected by light bending and, ultimately, to variations in the shape of the pulse profile \citep{Falkner+2018}. The flux resolved analysis (Fig.~\ref{fig:flux-resolved plots}) indicates that the overall shape of the pulse profile is similar at all flux levels -- excluding the brightest profiles only observed in Obs~2, which are discussed below. Except for these, the profiles appear also more similar across different observations at higher source flux.

In contrast, various accreting X-ray pulsars do show significant variations in their pulse shapes as a function of luminosity -- for recent examples see, e.g., \citet{Chhotaray+2024, Sharma:2024, Thalhammer:2024, DuYJ:2025}. In bright states, such changes are usually associated with the idea of passing a ``critical luminosity'' ($L_\mathrm{crit}$), above which the emitted radiation is capable of braking the flow above the surface of the neutron star, creating a radiative shock and an extended column \citep{Basko+Sunyaev:1976a, Becker:2012, Mushtukov:2015a}. At first order, $L_\mathrm{crit}$ depends on the mass accretion rate and the magnetic field strength of the source. Following \citet{Becker:2012}, \citet{Diez+2022} estimated $L_\mathrm{crit}$ for different assumptions on accretion geometry and magnetic field strength, finding ranges from $\sim 0.13$--$0.15\times10^{37} \mathrm{erg} \mathrm{s}^{-1}$ to $\sim 5$--$6\times10^{37} \mathrm{erg} \mathrm{s}^{-1}$. 
Using a competing approach, \citet{Mushtukov:2015a} stated that $L_\mathrm{crit}$ was likely to be around 
$10^{37} \mathrm{erg} \mathrm{s}^{-1}$. \citet{Diez+2022} obtained luminosities between $\sim 0.2$--$1\times10^{37} \mathrm{erg} \mathrm{s}^{-1}$ for one \nustar observation strictly simultaneous with Obs~3, and argued that Vela~X-1 lies between the sub and super-critical accretion regimes but that all such estimates have to be treated with caution. Based on the fluxes derived by \citet{Martinez+2014} for the same data as our brightest data points in Obs~2, we estimated an intrinsic luminosity of about $10^{37}$\,erg\,s$^{-1}$ for these. While the varying absorption does not allow to simply extrapolate this relation to the whole data set, all of the above indicates that the source was close to $L_\mathrm{crit}$ during the brightest pulse cycles analyzed in this work.

The brightest flux-resolved pulse profiles are different in that the amplitudes of $A$ and $B$ are almost identical (Fig.~\ref{fig:flux-resolved plots}). Thirty of the thirty-one pulse cycles that make up these profiles correspond to an almost continuous and relatively long time range (pulse cycles $\sim$ 218--267) preceding the flare of Obs~2, which was ignored in this work due to the disturbance of temporal information (see Sect.~\ref{sec:pulse periods and pulse cycles} for more details). \citet{Martinez+2014} found that the wind density started to increase during this observation at orbital phase $\sim 0.185$. Considering that they used $T_{\textrm{90}}$ as time of phase zero, and that we used $T_{\textrm{ecl}}$, their mentioned orbital phase corresponds to $\sim 0.160$ (pulse cycle 186) in our data. They claimed that an increase in the emission of the neutron star followed this wind enhancement, probably due to the corresponding gradual accretion of more mass. This also points to the source approaching $L_\mathrm{crit}$ during this period, which is reflected in the pulse profile as the change in relative amplitude between $A$ and $B$ and, ultimately, in the low PCC$_{8-10}$ values obtained for these pulse cycles (Fig.~\ref{fig:pcc-resolved plots}).

\subsection{Nonperiodic variations on short timescales}
\label{sec:nonperiodic variations at short timescales}

In the 8--10\,keV range, the majority of pulse cycles are, at most, $\sim 30$\% brighter or dimmer than their predecessor, while the most dramatic change was quantified by a factor of 2.25 (Obs~2). All of the above is in agreement with \citet{Staubert+1980}, who found smooth pulse-to-pulse intensity variation by up to a factor of 2. 

Concerning the subcomponents, $B1$ and $B2$ are less prone to vary overall, while $A1$ shows the largest variations. In terms of consecutive pulse cycles, $A2$ seems to vary less than the rest and $B2$ and $B3$ are more prone to fluctuate. It is interesting to note that the standard deviation of the flux fraction in $B2$ is one of the lowest and, at the same time, it fluctuates slightly more than the rest of sections from one pulse cycle to the next.

As found through the PCC-resolved analysis (Sect.~\ref{sec:pcc-resolved pulse profiles}, Fig.~\ref{fig:pcc-resolved plots}), the minority of pulse cycles with relatively low PCC tend to have $B$ (especially $B2$ and $B3$) relatively more prominent and $A$ somewhat suppressed, compared to mean profiles. At the same time, the HR shape does not change with PCC. If we assume that $A$ and $B$ are related to the two magnetic pole regions, variations in relative brightness imply variations in the relative accretion to the two poles. These abnormalities could be, e.g., sudden and slight increases (or decreases) in the density of the accreted matter. 

Overall, pulse-to-pulse variations in Vela X-1 are very stochastic, apparently following no temporal patterns nor trends. This is again in line with \citet{Staubert+1980}, but also with \citet{Frontera+1985}, who found ``chaotic spiky'' pulse shape variations on short timescales in the accreting X-ray pulsar 1A 0535+262.

\section{Summary and outlook}
\label{sec:summary and outlook}

We analyzed three long \xmmnewton observations from different years and orbital phases to study pulse profiles and pulse-to-pulse variations in the accreting X-ray pulsar Vela X-1.

We confirm that the overall shape of the pulse profile is stable, although we detected non-negligible differences on all timescales. The magnetic field and X-ray emission geometry appear very stable, and Vela X-1 seems to be accreting continuously but somewhat erratically from the stellar wind, leading to flux and pulse shape variations. The general structure of Vela X-1's pulse profile does not heavily depend on the flux.  
We also observed that, as the brightness increases, the similarity between profiles of different observations also rises. Below $\sim 3$\,keV, pulsations can be strongly suppressed by the variable strong absorption common in Vela X-1 at certain orbital phases. Variations at higher energies seem to be caused by changes in the density of the accretion flow. On short timescales, the pulse shape is also more stable when the source is brighter and variations are nonperiodic. There is no compelling evidence to suggest that any pulsed phase section varies significantly more than others.

It would be interesting to compare these profiles with new ones obtained from any future observations of Vela X-1 at different orbital phases. This would allow us to further study pulse profile variations along the binary orbit. More detailed models of the pulsed emission of Vela X-1 at soft and hard X-rays, and accounting for flux variations, could allow for a direct comparison with the observed profiles and provide additional insights into the causes of the observed variations. Extending this analysis to other accreting X-ray pulsars with complex pulse profiles would help us to better understand the variations presented here.

\begin{acknowledgements}
We thank the referee for very positive and constructive comments that identified some remaining gaps in the presented work, and led us to several additions to the text of this publication. VMF acknowledges ESA/ESAC and Universidad Complutense de Madrid (UCM) for the agreement that allows students from UCM to carry out their master’s thesis at ESAC, as this work started within the framework of this agreement. IEM acknowledges support from the grant ANID FONDECYT 11240206. SMN acknowledges support from grant PID2021-122955OB-C41 funded by MICIU/AEI/10.13039/501100011033 and by “ERDF/EU”. This research was supported by the International Space Science Institute (ISSI) in Bern, through ISSI International Team project $\#495$ (Feeding the spinning top). Work on this project profited from activities funded by the Space Science Faculty of the European Space Agency.  
\end{acknowledgements}

\bibliographystyle{aa}
\bibliography{references}

\begin{thebibliography}{49}
\expandafter\ifx\csname natexlab\endcsname\relax\def\natexlab#1{#1}\fi

\bibitem[{{Abalo} {et~al.}(2024){Abalo}, {Kretschmar}, {F{\"u}rst}, {Diez}, {El
  Mellah}, {Grinberg}, {Guainazzi}, {Mart{\'\i}nez-N{\'u}{\~n}ez},
  {Manousakis}, {Amato}, {Zhou}, \& {Beijersbergen}}]{Abalo+2024}
{Abalo}, L., {Kretschmar}, P., {F{\"u}rst}, F., {et~al.} 2024, \aap, 692, A188

\bibitem[{{Alonso-Hern{\'a}ndez} {et~al.}(2022){Alonso-Hern{\'a}ndez},
  {F{\"u}rst}, {Kretschmar}, {Caballero}, \& {Joyce}}]{Alonso+2022}
{Alonso-Hern{\'a}ndez}, J., {F{\"u}rst}, F., {Kretschmar}, P., {Caballero}, I.,
  \& {Joyce}, A.~M. 2022, \aap, 662, A62

\bibitem[{{Astropy Collaboration} {et~al.}(2013){Astropy Collaboration},
  {Robitaille}, {Tollerud}, {Greenfield}, {Droettboom}, {Bray}, {Aldcroft},
  {Davis}, {Ginsburg}, {Price-Whelan}, {Kerzendorf}, {Conley}, {Crighton},
  {Barbary}, {Muna}, {Ferguson}, {Grollier}, {Parikh}, {Nair}, {Unther},
  {Deil}, {Woillez}, {Conseil}, {Kramer}, {Turner}, {Singer}, {Fox}, {Weaver},
  {Zabalza}, {Edwards}, {Azalee Bostroem}, {Burke}, {Casey}, {Crawford},
  {Dencheva}, {Ely}, {Jenness}, {Labrie}, {Lim}, {Pierfederici}, {Pontzen},
  {Ptak}, {Refsdal}, {Servillat}, \& {Streicher}}]{Astropycollab+2013}
{Astropy Collaboration}, {Robitaille}, T.~P., {Tollerud}, E.~J., {et~al.} 2013,
  \aap, 558, A33

\bibitem[{{Bachetti} {et~al.}(2024){Bachetti}, {Huppenkothen}, {Stevens},
  {Swinbank}, {Mastroserio}, {Lucchini}, {Lai}, {Buchner}, {Desai}, {Joshi},
  {Pisanu}, {Pisupati}, {Sharma}, {Tripathi}, \& {Vats}}]{Bachetti+2024}
{Bachetti}, M., {Huppenkothen}, D., {Stevens}, A., {et~al.} 2024, The Journal
  of Open Source Software, 9, 7389

\bibitem[{{Basko} \& {Sunyaev}(1976)}]{Basko+Sunyaev:1976a}
{Basko}, M.~M. \& {Sunyaev}, R.~A. 1976, \mnras, 175, 395

\bibitem[{{Becker} {et~al.}(2012){Becker}, {Klochkov}, {Sch{\"o}nherr},
  {Nishimura}, {Ferrigno}, {Caballero}, {Kretschmar}, {Wolff}, {Wilms}, \&
  {Staubert}}]{Becker:2012}
{Becker}, P.~A., {Klochkov}, D., {Sch{\"o}nherr}, G., {et~al.} 2012, \aap, 544,
  A123

\bibitem[{{Boynton} {et~al.}(1984){Boynton}, {Deeter}, {Lamb}, {Zylstra},
  {Pravdo}, {White}, {Wood}, \& {Yentis}}]{Boynton+1984}
{Boynton}, P.~E., {Deeter}, J.~E., {Lamb}, F.~K., {et~al.} 1984, \apjl, 283,
  L53

\bibitem[{{Bulik} {et~al.}(1995){Bulik}, {Riffert}, {Meszaros}, {Makishima},
  {Mihara}, \& {Thomas}}]{Bulik+1995}
{Bulik}, T., {Riffert}, H., {Meszaros}, P., {et~al.} 1995, \apj, 444, 405

\bibitem[{{Chhotaray} {et~al.}(2024){Chhotaray}, {Jaisawal}, {Nandi}, {Naik},
  {Kumari}, {Ng}, \& {Gendreau}}]{Chhotaray+2024}
{Chhotaray}, B., {Jaisawal}, G.~K., {Nandi}, P., {et~al.} 2024, \apj, 963, 132

\bibitem[{{Chodil} {et~al.}(1967){Chodil}, {Mark}, {Rodrigues}, {Seward}, \&
  {Swift}}]{Chodil+1967}
{Chodil}, G., {Mark}, H., {Rodrigues}, R., {Seward}, F.~D., \& {Swift}, C.~D.
  1967, \apj, 150, 57

\bibitem[{{Deeter} {et~al.}(1989){Deeter}, {Boynton}, {Lamb}, \&
  {Zylstra}}]{Deeter+1989}
{Deeter}, J.~E., {Boynton}, P.~E., {Lamb}, F.~K., \& {Zylstra}, G. 1989, \apj,
  336, 376

\bibitem[{{Diez} {et~al.}(2023){Diez}, {Grinberg}, {F{\"u}rst}, {El Mellah},
  {Zhou}, {Santangelo}, {Mart{\'\i}nez-N{\'u}{\~n}ez}, {Amato}, {Hell}, \&
  {Kretschmar}}]{Diez+2023}
{Diez}, C.~M., {Grinberg}, V., {F{\"u}rst}, F., {et~al.} 2023, \aap, 674, A147

\bibitem[{{Diez} {et~al.}(2022){Diez}, {Grinberg}, {F{\"u}rst},
  {Sokolova-Lapa}, {Santangelo}, {Wilms}, {Pottschmidt},
  {Mart{\'\i}nez-N{\'u}{\~n}ez}, {Malacaria}, \& {Kretschmar}}]{Diez+2022}
{Diez}, C.~M., {Grinberg}, V., {F{\"u}rst}, F., {et~al.} 2022, \aap, 660, A19

\bibitem[{{Doroshenko} {et~al.}(2013){Doroshenko}, {Santangelo}, {Nakahira},
  {Mihara}, {Sugizaki}, {Matsuoka}, {Nakajima}, \&
  {Makishima}}]{Doroshenko:2013}
{Doroshenko}, V., {Santangelo}, A., {Nakahira}, S., {et~al.} 2013, \aap, 554,
  A37

\bibitem[{{Du} {et~al.}(2025){Du}, {Ducci}, {Ji}, {Bu}, {Kong}, {Wang}, {Tuo},
  \& {Santangelo}}]{DuYJ:2025}
{Du}, Y.-J., {Ducci}, L., {Ji}, L., {et~al.} 2025, \aap, 694, A156

\bibitem[{{Falkner}(2018)}]{Falkner+2018}
{Falkner}, S. 2018, PhD thesis, Friedrich Alexander University of
  Erlangen-Nuremberg, Germany

\bibitem[{{Frontera} {et~al.}(1985){Frontera}, {dal Fiume}, {Morelli}, \&
  {Spada}}]{Frontera+1985}
{Frontera}, F., {dal Fiume}, D., {Morelli}, E., \& {Spada}, G. 1985, \apj, 298,
  585

\bibitem[{{F{\"u}rst} {et~al.}(2010){F{\"u}rst}, {Kreykenbohm}, {Pottschmidt},
  {Wilms}, {Hanke}, {Rothschild}, {Kretschmar}, {Schulz}, {Huenemoerder},
  {Klochkov}, \& {Staubert}}]{Fuerst:2010}
{F{\"u}rst}, F., {Kreykenbohm}, I., {Pottschmidt}, K., {et~al.} 2010, \aap,
  519, A37+

\bibitem[{{F{\"u}rst} {et~al.}(2014){F{\"u}rst}, {Pottschmidt}, {Wilms},
  {Tomsick}, {Bachetti}, {Boggs}, {Christensen}, {Craig}, {Grefenstette},
  {Hailey}, {Harrison}, {Madsen}, {Miller}, {Stern}, {Walton}, \&
  {Zhang}}]{Furst+2014}
{F{\"u}rst}, F., {Pottschmidt}, K., {Wilms}, J., {et~al.} 2014, \apj, 780, 133

\bibitem[{{F{\"u}rst} {et~al.}(2011){F{\"u}rst}, {Suchy}, {Kreykenbohm},
  {Barrag{\'a}n}, {Wilms}, {Pottschmidt}, {Caballero}, {Kretschmar},
  {Ferrigno}, \& {Rothschild}}]{Furst+2011}
{F{\"u}rst}, F., {Suchy}, S., {Kreykenbohm}, I., {et~al.} 2011, \aap, 535, A9

\bibitem[{{Houck} \& {Denicola}(2000)}]{Houck+2000}
{Houck}, J.~C. \& {Denicola}, L.~A. 2000, in Astronomical Society of the
  Pacific Conference Series, Vol. 216, Astronomical Data Analysis Software and
  Systems IX, ed. N.~{Manset}, C.~{Veillet}, \& D.~{Crabtree}, 591

\bibitem[{{Huppenkothen} {et~al.}(2019){Huppenkothen}, {Bachetti}, {Stevens},
  {Migliari}, {Balm}, {Hammad}, {Khan}, {Mishra}, {Rashid}, {Sharma}, {Martinez
  Ribeiro}, \& {Valles Blanco}}]{Huppenkothen+2019}
{Huppenkothen}, D., {Bachetti}, M., {Stevens}, A.~L., {et~al.} 2019, \apj, 881,
  39

\bibitem[{{Kaper} {et~al.}(1994){Kaper}, {Hammerschlag-Hensberge}, \&
  {Zuiderwijk}}]{Kaper+1994}
{Kaper}, L., {Hammerschlag-Hensberge}, G., \& {Zuiderwijk}, E.~J. 1994, \aap,
  289, 846

\bibitem[{{Klochkov} {et~al.}(2011){Klochkov}, {Staubert}, {Santangelo},
  {Rothschild}, \& {Ferrigno}}]{Klochkov+2011}
{Klochkov}, D., {Staubert}, R., {Santangelo}, A., {Rothschild}, R.~E., \&
  {Ferrigno}, C. 2011, \aap, 532, A126

\bibitem[{{Kraus} {et~al.}(1996){Kraus}, {Blum}, {Schulte}, {Ruder}, \&
  {Meszaros}}]{Kraus+1986}
{Kraus}, U., {Blum}, S., {Schulte}, J., {Ruder}, H., \& {Meszaros}, P. 1996,
  \apj, 467, 794

\bibitem[{{Kretschmar} {et~al.}(2021){Kretschmar}, {El Mellah},
  {Mart{\'\i}nez-N{\'u}{\~n}ez}, {F{\"u}rst}, {Grinberg}, {Sander}, {van den
  Eijnden}, {Degenaar}, {Ma{\'\i}z Apell{\'a}niz}, {Jim{\'e}nez Esteban},
  {Ramos-Lerate}, \& {Utrilla}}]{Kretschmar+2021}
{Kretschmar}, P., {El Mellah}, I., {Mart{\'\i}nez-N{\'u}{\~n}ez}, S., {et~al.}
  2021, \aap, 652, A95

\bibitem[{{Kretschmar} {et~al.}(2014){Kretschmar}, {Marcu}, {K{\"u}hnel},
  {Klochkov}, {Pottschmidt}, {Staubert}, {Wilson-Hodge}, {Jenke}, {Caballero},
  \& {F{\"u}rst}}]{Kretschmar+2014}
{Kretschmar}, P., {Marcu}, D., {K{\"u}hnel}, M., {et~al.} 2014, in European
  Physical Journal Web of Conferences, Vol.~64, European Physical Journal Web
  of Conferences (EDP), 06012

\bibitem[{{Kreykenbohm} {et~al.}(2008){Kreykenbohm}, {Wilms}, {Kretschmar},
  {Torrej{\'o}n}, {Pottschmidt}, {Hanke}, {Santangelo}, {Ferrigno}, \&
  {Staubert}}]{Kreykenbohm+2008}
{Kreykenbohm}, I., {Wilms}, J., {Kretschmar}, P., {et~al.} 2008, \aap, 492, 511

\bibitem[{{La Barbera} {et~al.}(2003){La Barbera}, {Santangelo}, {Orlandini},
  \& {Segreto}}]{Barbera+2003}
{La Barbera}, A., {Santangelo}, A., {Orlandini}, M., \& {Segreto}, A. 2003,
  \aap, 400, 993

\bibitem[{{Laycock} {et~al.}(2025){Laycock}, {Cappallo}, {Pradhan},
  {Christodoulou}, \& {Paul}}]{Laycock:2025}
{Laycock}, S. G.~T., {Cappallo}, R.~C., {Pradhan}, P., {Christodoulou}, D.~M.,
  \& {Paul}, B. 2025, \apj, 978, 80

\bibitem[{{Leahy} \& {Li}(1995)}]{Leahy+Li:1995}
{Leahy}, D.~A. \& {Li}, L. 1995, \mnras, 277, 1177

\bibitem[{{Lomaeva} {et~al.}(2020){Lomaeva}, {Grinberg}, {Guainazzi}, {Hell},
  {Bianchi}, {Bissinger n{\'e} K{\"u}hnel}, {F{\"u}rst}, {Kretschmar},
  {Mart{\'\i}nez-Chicharro}, {Mart{\'\i}nez-N{\'u}{\~n}ez}, \&
  {Torrej{\'o}n}}]{Lomaeva+2020}
{Lomaeva}, M., {Grinberg}, V., {Guainazzi}, M., {et~al.} 2020, \aap, 641, A144

\bibitem[{{Malacaria} {et~al.}(2016){Malacaria}, {Mihara}, {Santangelo},
  {Makishima}, {Matsuoka}, {Morii}, \& {Sugizaki}}]{Malacaria:2016}
{Malacaria}, C., {Mihara}, T., {Santangelo}, A., {et~al.} 2016, \aap, 588, A100

\bibitem[{{Martin-Carrillo} {et~al.}(2012){Martin-Carrillo}, {Kirsch},
  {Caballero}, {Freyberg}, {Ibarra}, {Kendziorra}, {Lammers}, {Mukerjee},
  {Sch{\"o}nherr}, {Stuhlinger}, {Saxton}, {Staubert}, {Suchy}, {Wellbrock},
  {Webb}, \& {Guainazzi}}]{MartinCarrillo+2012}
{Martin-Carrillo}, A., {Kirsch}, M.~G.~F., {Caballero}, I., {et~al.} 2012,
  \aap, 545, A126

\bibitem[{{Mart{\'\i}nez-N{\'u}{\~n}ez}
  {et~al.}(2014){Mart{\'\i}nez-N{\'u}{\~n}ez}, {Torrej{\'o}n}, {K{\"u}hnel},
  {Kretschmar}, {Stuhlinger}, {Rodes-Roca}, {F{\"u}rst}, {Kreykenbohm},
  {Martin-Carrillo}, {Pollock}, \& {Wilms}}]{Martinez+2014}
{Mart{\'\i}nez-N{\'u}{\~n}ez}, S., {Torrej{\'o}n}, J.~M., {K{\"u}hnel}, M.,
  {et~al.} 2014, \aap, 563, A70

\bibitem[{{Maxwell}(2011)}]{Maxwell+2011}
{Maxwell}, E.~A. 2011, arXiv e-prints, arXiv:1102.0822

\bibitem[{{McClintock} {et~al.}(1976){McClintock}, {Rappaport}, {Joss},
  {Bradt}, {Buff}, {Clark}, {Hearn}, {Lewin}, {Matilsky}, {Mayer}, \&
  {Primini}}]{McClintock+1976}
{McClintock}, J.~E., {Rappaport}, S., {Joss}, P.~C., {et~al.} 1976, \apjl, 206,
  L99

\bibitem[{{M{\"u}ller} {et~al.}(2011){M{\"u}ller}, {Klochkov}, {Santangelo},
  {Mihara}, \& {Sugizaki}}]{Mueller+2011}
{M{\"u}ller}, D., {Klochkov}, D., {Santangelo}, A., {Mihara}, T., \&
  {Sugizaki}, M. 2011, \aap, 535, A102

\bibitem[{{Mushtukov} {et~al.}(2015){Mushtukov}, {Suleimanov}, {Tsygankov}, \&
  {Poutanen}}]{Mushtukov:2015a}
{Mushtukov}, A.~A., {Suleimanov}, V.~F., {Tsygankov}, S.~S., \& {Poutanen}, J.
  2015, \mnras, 447, 1847

\bibitem[{{Raubenheimer}(1990)}]{Raubenheimer+1990}
{Raubenheimer}, B.~C. 1990, \aap, 234, 172

\bibitem[{{Sanjurjo-Ferr{\'\i}n} {et~al.}(2025){Sanjurjo-Ferr{\'\i}n},
  {Torrej{\'o}n}, {Postnov}, {Nowak}, {Rodes-Roca}, {Oskinova},
  {Planelles-Villalva}, \& {Schulz}}]{Sanjurjo-Ferrin:2025}
{Sanjurjo-Ferr{\'\i}n}, G., {Torrej{\'o}n}, J.~M., {Postnov}, K., {et~al.}
  2025, \aap, 694, A192

\bibitem[{{Sato} {et~al.}(1986){Sato}, {Hayakawa}, {Nagase}, {Masai}, {Dotani},
  {Inoue}, {Makino}, {Makishima}, \& {Ohashi}}]{Sato:1986PASJ}
{Sato}, N., {Hayakawa}, S., {Nagase}, F., {et~al.} 1986, \pasj, 38, 731

\bibitem[{{Sharma} {et~al.}(2024){Sharma}, {Mandal}, {Pal}, {Paul}, {Jaisawal},
  \& {Ratheesh}}]{Sharma:2024}
{Sharma}, R., {Mandal}, M., {Pal}, S., {et~al.} 2024, \mnras, 534, 1028

\bibitem[{{Staubert} {et~al.}(1980){Staubert}, {Kendziorra}, {Pietsch},
  {Reppin}, {Truemper}, \& {Voges}}]{Staubert+1980}
{Staubert}, R., {Kendziorra}, E., {Pietsch}, W., {et~al.} 1980, \apj, 239, 1010

\bibitem[{{Staubert} {et~al.}(2004){Staubert}, {Kreykenbohm}, {Kretschmar},
  {Chernyakova}, {Pottschmidt}, {Benlloch-Garcia}, {Wilms}, {Santangelo},
  {Segreto}, {von Kienlin}, {Sidoli}, {Larsson}, \&
  {Westergaard}}]{Staubert+2004}
{Staubert}, R., {Kreykenbohm}, I., {Kretschmar}, P., {et~al.} 2004, in ESA
  Special Publication, Vol. 552, 5th INTEGRAL Workshop on the INTEGRAL
  Universe, ed. V.~{Schoenfelder}, G.~{Lichti}, \& C.~{Winkler}, 259

\bibitem[{{Str{\"u}der} {et~al.}(2001){Str{\"u}der}, {Briel}, {Dennerl},
  {Hartmann}, {Kendziorra}, {Meidinger}, {Pfeffermann}, {Reppin}, {Aschenbach},
  {Bornemann}, {Br{\"a}uninger}, {Burkert}, {Elender}, {Freyberg}, {Haberl},
  {Hartner}, {Heuschmann}, {Hippmann}, {Kastelic}, {Kemmer}, {Kettenring},
  {Kink}, {Krause}, {M{\"u}ller}, {Oppitz}, {Pietsch}, {Popp}, {Predehl},
  {Read}, {Stephan}, {St{\"o}tter}, {Tr{\"u}mper}, {Holl}, {Kemmer}, {Soltau},
  {St{\"o}tter}, {Weber}, {Weichert}, {von Zanthier}, {Carathanassis}, {Lutz},
  {Richter}, {Solc}, {B{\"o}ttcher}, {Kuster}, {Staubert}, {Abbey}, {Holland},
  {Turner}, {Balasini}, {Bignami}, {La Palombara}, {Villa}, {Buttler},
  {Gianini}, {Lain{\'e}}, {Lumb}, \& {Dhez}}]{Strueder+2001}
{Str{\"u}der}, L., {Briel}, U., {Dennerl}, K., {et~al.} 2001, \aap, 365, L18

\bibitem[{{Sturner} \& {Dermer}(1994)}]{Sturner+Dermer:1994}
{Sturner}, S.~J. \& {Dermer}, C.~D. 1994, \aap, 284, 161

\bibitem[{{Thalhammer} {et~al.}(2024){Thalhammer}, {Ballhausen},
  {Sokolova-Lapa}, {Stierhof}, {Zainab}, {Staubert}, {Pottschmidt}, {Coley},
  {Rothschild}, {Jaisawal}, {West}, {Becker}, {Pradhan}, {Kretschmar}, \&
  {Wilms}}]{Thalhammer:2024}
{Thalhammer}, P., {Ballhausen}, R., {Sokolova-Lapa}, E., {et~al.} 2024, \aap,
  688, A213

\bibitem[{{Virtanen} {et~al.}(2020){Virtanen}, {Gommers}, {Oliphant},
  {Haberland}, {Reddy}, {Cournapeau}, {Burovski}, {Peterson}, {Weckesser},
  {Bright}, {van der Walt}, {Brett}, {Wilson}, {Millman}, {Mayorov}, {Nelson},
  {Jones}, {Kern}, {Larson}, {Carey}, {Polat}, {Feng}, {Moore}, {VanderPlas},
  {Laxalde}, {Perktold}, {Cimrman}, {Henriksen}, {Quintero}, {Harris},
  {Archibald}, {Ribeiro}, {Pedregosa}, {van Mulbregt}, \& {SciPy 1. 0
  Contributors}}]{Virtanen+2020}
{Virtanen}, P., {Gommers}, R., {Oliphant}, T.~E., {et~al.} 2020, Nature
  Methods, 17, 261

\end{thebibliography}

\end{document}